\documentclass[a4paper,10pt]{article}
\usepackage[X2,T2A]{fontenc}
\usepackage[cp1251]{inputenc}
\usepackage[bulgarian]{babel}

\usepackage{amssymb}
\usepackage{eurosym}
\usepackage{longtable}
\usepackage{booktabs}

\usepackage{sectsty}
\allsectionsfont{\sffamily}

\newcommand{\Hat}{{\usefont{X2}{cmr}{m}{n}\^{}}}

\newcommand{\ar}{$\rightarrow$}
\newcommand{\tabtit}[1]{\textit{\textbf{#1}}}

\newcommand{\key}[1]{\multicolumn{3}{|l|}{#1}}
\newcommand{\keyn}[1]{\multicolumn{3}{|l|}{}}
\newcommand{\keys}[1]{\multicolumn{3}{|l|}{\scriptsize{#1}}}
\newcommand{\keyspace}[2]{\multicolumn{#1}{|l|}{#2}}
\newcommand{\keyname}[1]{{\scriptsize { }}  }

\newenvironment{keyboard}{
\begin{longtable}{@{}c@{}c@{}c@{}c@{}c@{}c@{}c@{}c@{}c@{}c@{}c@{}c@{}c@{}c@{}c@{}c@{}c@{}c@{}c@{}c@{}c@{}c@{}c@{}c@{}c@{}c@{}c@{}c@{}c@{}c@{}c@{}c@{}c@{}c@{}c@{}c@{}c@{}c@{}c@{}c@{}c@{}c@{}c@{}c@{}}

\hspace{8pt} & \hspace{8pt} & \hspace{8pt} & \hspace{8pt} & \hspace{8pt} & \hspace{8pt} & \hspace{8pt} & \hspace{8pt} & \hspace{8pt} & \hspace{8pt} & \hspace{8pt} & \hspace{8pt} & \hspace{8pt} & \hspace{8pt} & \hspace{8pt} & \hspace{8pt} & \hspace{8pt} & \hspace{8pt} & \hspace{8pt} & \hspace{8pt} & \hspace{8pt} & \hspace{8pt} & \hspace{8pt} & \hspace{8pt} & \hspace{8pt} & \hspace{8pt} & \hspace{8pt} & \hspace{8pt} & \hspace{8pt} & \hspace{8pt} & \hspace{8pt} & \hspace{8pt} & \hspace{8pt} & \hspace{8pt} & \hspace{8pt} & \hspace{8pt} & \hspace{8pt} & \hspace{8pt} & \hspace{8pt} & \hspace{8pt} & \hspace{8pt} & \hspace{8pt} & \hspace{8pt} & \hspace{8pt}  \\
}
{
\end{longtable}
}

\newcommand{\subsymbol}[3]{
  \multicolumn{1}{||c||}{\makebox[4.2cm]{\small{#1}}} &
  \multicolumn{1}{c||}{\makebox[3.5cm]{\small{#2}}} &
  \multicolumn{1}{c||}{\makebox[3.4cm]{\small{#3}}} \\
  \hline
}

\newenvironment{substable}{
  ~\newline
  \begin{tabular}{ccc}
    {\scriptsize \textbf{\textit{знакът}:}} & 
    {\scriptsize \textbf{\textit{се поставя при}:}} &
    {\scriptsize \textbf{\textit{причина}:}} \\
    \hline
}{
  \end{tabular}
  \vspace{2mm}
}

\begin{document}

\title{Разширени български клавиатурни подредби
\thanks{
  Този документ е резултат от публично обсъждане на това какви трябва
  да бъдат българските клавиатури при ГНУ/Линукс.  Първата му версия
  бе публикувана на 5~август 2005~г. и в настоящия си вид той не би
  съществувал, ако не бяха многобройните препоръки, които авторът
  получи оттогава.  Въпреки че за всички недостатъци в предложените
  клавиатурни подредби, които читателят би намерил, авторът поема
  пълна отговорност, той би искал да изкаже по-специална благодарност
  на Александър Шопов и Огнян Кулев за ползотворното обсъждане какви
  знаци трябва да се поддържат, на Валентин Стойков и Ради Пипев за
  предложените от тях варианти за разположението на знаците от трети и
  четвърти регистър, на Михаил Балабанов за някои сведения относно
  клавиатурните възможности на MS Windows, на Дамян Иванов за
  „моралната“ подкрепа, която той оказа на много критикуваната нова
  фонетична клавиатура в БДС 5237:2006, както и на авторите на новия
  стандарт "--- Димитър Добрев и проф.~Димитър Скордев.
}}

\author{Антон Зиновиев}

\date{17 февруари 2007}

\maketitle

\begin{abstract}
  Старият български клавиатурен стандарт БДС 5237-78 бе изработен за
  използване предимно при пишещи машини.  Масовото разпространение на
  електронно-изчислителната техника наложи неговото осъвременяване.
  От една страна това е свързано с необходимистта да се поддържат
  знаци като българските кавички, буквата „и със знак във вид на
  ударение“, дълго тире и др.  От друга страна широко разпространене
  получи и т.н. фонетична клавиатура, която не бе стандартизирана по
  БДС.

  В този документ е направен анализ на възможностите за подобряване
  както на клавиатурата по БДС 5237-78, така и на традиционната
  фонетична клавиатура.  При това е направено сравнение и с новия
  български клавиатурен стандарт БДС 5237:2006.  Предложен е алгоритъм
  за разпределение на допълнителни знаци върху клавишите на
  клавиатурата, така че те да бъдат лесно намирани от потребителите
  дори и да не са надписани върху клавишите.

  Дадени са формални дефиниции и са илюстрирани четири клавиатурни
  подредби "--- три разширени клавиатурни подредби за писане в режим
  „кирилица“ (тип „БДС“, фонетичен тип и фонетичен тип според БДС
  5237:2006) и една разширена клавиатурна подредба за писане в режим
  „латиница“.
\end{abstract}

\tableofcontents

\section{Предговор}

Дълго време пишещите машини в България не са имали стандартно
разположение на знаците върху клавишите.  Чак през 1907~г. по
инициатива на стенографите, работещи в Народното събрание,
безпорядъкът се отстранява като въз основа на статистически анализ на
разнородни текстове се съставя проектът за българските знаци на
клавиатурите на пишещите машини, известен днес като клавиатура по
„БДС“.

От съставянето си до днес, този стандарт претърпява известни
изменения.  Първото се състои в премахването на клавишите за ненужните
вече букви „ят“ ({\usefont{X2}{cmr}{m}{n}\cyryat}) и „голям юс“
({\usefont{X2}{cmr}{m}{n}\cyrbyus}) и замяната им с руските букви „ы“
и~„э“.  През 1978~г. се приема втора поправка в стандарта, касаеща
местата на голямото тире, на знака за номер (№) и на римските цифри I
и~V.

При писането на пишеща машина не е възможно да се постигне полиграфско
качество на отпечатания текст.  От една страна това е невъзможно,
защото пишещите машини използват шрифт, при който всички знаци имат
равна ширина.  От друга страна много от знаците просто липсват на
клавиатурата, което налага при нужда те да бъдат дописвани на ръка или
да бъде използвано тяхно приближение измежду наличните знаци.  Така
например обичайна практика е да се използва буквата „и кратко“ (й)
вместо удареното и (\`и) и общ знак за кавички (") вместо отделни
знаци за отварящите („) и затварящите (“) кавички\footnote{Някои
  интелигентни текстообработващи програми заменят автоматично
  неутралните кавички (") с двустранни кавички.  Доколкото е известно
  на автора, измежду тях единствено OpenOffice.org поставя правилни
  български кавички.  Наличните до момента версии на MS Word поставят
  или американски кавички (по-старите версии), или неправилни
  затварящи кавички (по-новите версии).  Разбира се, не можем да
  очакваме повечето програми да заменят автоматично неутралните
  кавички с двустранни, тъй като в много случаи тази „интелигентна“
  замяна е нежелателна.}.

През декември 2006~г. в Института по стандартизация бе приет български
държавен клавиатурен стандарт.  В него досегашната стандартна
клавиатурна подредба бе модернизирана, като бе добавена поддръжка за
българските кавички, дългото тире, удареното „и“ и знаците за евро и
долар.  Освен това този стандарт описва и фонетична клавиатурна
подредба.  Разположението на буквите при нея обаче се различава
значително от това на досега използваната фонетична клавиатура.

С масовото навлизане на електронно-изчислителна техника възникнаха
нови изисквания по отношение на компютърните клавиатури.  Шрифтовете,
използвани от компютрите, поддържат огромно количество различни знаци
и често липсата на даден знак върху клавиатурата е единствената
причина потребителят да не го използва.  От друга страна за разлика от
повечето пишещи машини, при компютрите обикновено не е проблем
клавиатурните клавиши да са в състояние да генерират до осем различни
знаци "--- четири в режим „кирилица“ и четири в режим „латиница“.

В този документ е описан разширен вариант на стандартната латинска
клавиатурна подредба QWERTY, при който е добавена поддръжка за
разнообразни препинателни, математически, технически и др. знаци като
{\usefont{T1}{cmr}{m}{n}\pounds}, $\Sigma$, $\leqq$, $\approx$, $\pm$
и~др.

Освен това тук е описан и разширен вариант на клавиатурната подредба
тип „БДС“, както и две разширени фонетични подредби "--- една, съобразно
новия стандарт БДС 5237:2006, и друга с обичайното разположение на
буквите.

\section{Дефиниции и ограничения}

Този документ не дефинира поведението и разположението на служебните
клавиши, които не генерират графични знаци.  Такива клавиши са
например клавишът за интервал и клавиши като Shift, Control, Alt,
AltGr, Caps Lock, Tab, Esc, Insert, PageUp, Command, Compose, Menu,
Option, Power и др.

Не се дефинира и поведението на клавишите, принадлежащи на цифрови
клавиатури, т.е. на клавиатури, чието основно предназначение е
въвеждането на числа.  Не се определя и поведението на клавишите от
цифровия блок в дясната част на компютърните
клавиатури\footnote{Според БДС 5237:2006 клавишът от цифровия блок
  обозначен с „Del“ при активен NumLock трябва да генерира десетична
  запетая, а не десетична точка.  Тъй като този стандарт не определя
  поведението на клавиатурата в режим „латиница“, вероятно трябва да
  се подразбира, че в режим „латиница“ същият клавиш генерира
  десетична точка.}.

Компютърната клавиатура трябва да може да се превключва в два режима.
В тези два режима клавишите на клавиатурата генерират различен набор
знаци.  Единият от двата режима се използва за писане с латински букви
и ще го наричаме режим „латиница“, а другият се използва за писане с
кирилски букви и ще го наричаме режим „кирилица“.

Писменият знак, който даден клавиш генерира, когато клавиатурата се
намира в обичайното си състояние в кой да е от тези два режима ще
наричаме знак от първи регистър.

Писменият знак, който даден клавиш генерира, когато е натиснат
клавишът Shift (при неактивен превключвател CapsLock), ще наричаме
знак от втори регистър.

Клавишите на клавиатурата може (но не е задължително) да са в
състояние да генерират още два допълнителни знака "--- ще ги наричаме
знаци от трети и четвърти регистър.

За краткост вместо за клавиатурни подредби тип „БДС“, фонетична
клавиатурна подредба и фонетична клавиатурна подредба по БДС 5237:2006
ще говорим за клавиатура тип „БДС“, фонетична клавиатура и фонетична
клавиатура по БДС.

Клавишът, който в режим „латиница“ генерира знаците $\backslash$ и~$|$
и се намира в дясната част на клавиатурата в близост до клавиша
„Enter“, ще наричаме клавиш~<BKSL>.  Някои клавиатури имат и един
допълнителен буквено-цифров клавиш, който се намира между левия клавиш
„Shift“ и клавиша, който в режим „латиница“ генерира буквата „Z“.
Този клавиш ще наричаме клавиш~<LSGT>.

Вместо „буква и с надреден знак във вид на ударение“ ще пишем просто
„и с ударение“.

\section{Обсъждане на знаците от първи и втори регистър}

Тъй като клавиатурата не винаги е в състояние на генерира знаци от
трети и четвърти регистър, е важно най-важните знаци да са разположени
в първи и втори регистър.  Дори и да има възможност да се използват
знаци от трети и четвърти регистър, достъпът до тези знаци често е
неудобен и затова е важно най-употребяваните знаци да бъдат
разположени в първи или втори регистър.







\subsection{Традиционна клавиатура тип „БДС“}\label{sec:bds12}

Стриктната клавиатурна подредба според БДС 5237-78 не поддържа
следните необходими писмени знаци: кавички („“), тире (--), ударено и
(\`{и}), евро (\euro).  От друга страна тя притежава следния резерв от
осем знака, които не са особено полезни в режим „кирилица“:

\subsubsection*{Клавишът <BKSL> (два знака)}

Клавишът, който в режим „латиница“ генерира знаците $\backslash$ и $|$
се намира на различно място при различните клавиатури.  Според БДС
5237-78 този клавиш трябва да се намира в края на третия ред и в режим
„кирилица“ трябва да генерира кръгли скоби.  На повечето клавиатури
обаче скобите са нарисувани върху най-левия клавиш от първия ред, за
който БДС 5237-78 не казва нищо.

Това означава, че един от тези два клавиша е свободен.  Както според
новия стандарт БДС 5237:2006, така и според клавиатурата тип „БДС“,
описана в този документ, кръглите скоби трябва да се генерират от
най-левия клавиш от първия ред на клавиатурата.  Следователно клавишът
„<BKSL>“ остава е свободен.

Според БДС 5237:2006 този клавиш трябва да генерира българските
кавички.  Такава употреба на отделен клавиш за двата вида кавички
кавичките е аналогична с използването при тази клавиатурна подредба на
отделен клавиш за двата вида скоби "--- отваряща и затваряща.

\subsubsection*{Римските цифри „I“ и „V“ (два знака)}

Според стандарта БДС 5237-78 два от клавишите генерират в горен
регистър римските цифри за 1 и 5 (т.е. I и V).  Тъй като този стандарт
е предназначен за използване на пишещи машини, вместо римската цифра
за 10 (т.е. X) се предвижда въвеждането на главната кирилска буква Х.
При използването на компютри обаче е нежелателно римските числа да се
изписват смесено с латински (I и V) и кирилски (Х) букви.

Според БДС 5237:2006 на мястото на римските цифри I и V се поставят
съответно знаците за долар и евро.

\subsubsection*{Знак при клавиша за цифрата „9“ (един знак)}

Става въпрос за знака, който се генерира във втори регистър от клавиша
с цифрата „9“.  Според стандарта БДС 5237-78 не става ясно дали този
знак представлява дълго тире или знак за подчертаване.  Обикновено
пишещите машини и компютърните клавиатури генерират знака за
подчертаване.  Макар и да е полезен понякога при пишещите машини, при
компютрите този клавиш не може да се използва за подчертаване на текст
и присъствието му на клавиатурите е ненужно.

Според БДС 5237:2006 на мястото на знака за подчертаване се поставя
дълго тире.  Според българския правописен правилник употребата на
дълго и късо тире е строго регламентирана и това означава, че
клавиатурата трябва да поддържа и двете вида тирета.  Поставянето на
дългото тире на мястото на знака за подчертаване е естествено предвид
визуалното сходство на тези два знака.

\subsubsection*{Руските букви „ы“ и~„э“ (три знака)}

Тъй като правописната реформа от 1945~г. отменя употребата на буквите
„ят“ ({\usefont{X2}{cmr}{m}{n}\cyryat}) и „голям юс“
({\usefont{X2}{cmr}{m}{n}\cyrbyus}), е взето решение на пишещите
машини те да бъдат заменени с руските букви „ы“ и~„э“.  Днешните
обществено-политически условия не изискват задължителното присъствие
на тези две букви на всяка българска клавиатура и те могат да се
заменят с по-полезни и по-често употребявани знаци.

Според новия държавен клавиатурен стандарт руските букви остават на
българската клавиатура.  Според същия стандарт обаче, фонетичната
клавиатура няма да поддържа руските букви.  Няма абсолютно никакви
технически пречки руските букви да се добавят и при фонетичния вариант
за разположение на знаците върху клавишите.  Ако се прецени, че
руските букви са полезни, то те трябваше да присъстват и при
фонетичната клавиатура.  Ако пък обратно "--- прецени се, че руските
букви не са достатъчно полезни, то те би следвало да бяха заменени с
по-полезни от тях и при клавиатурата тип „БДС“.  Възможно е също тези
букви да са били оставени с цел да не се променя значително старата
клавиатурна подредба тип „БДС“.  В този случай обаче възниква въпросът
защо при клавиатурата тип „БДС“ са запазени не особено полезни знаци,
докато в същото време при фонетичната клавиатура буквите са претърпели
значителни размествания.  Ако е вярно първото, тогава новият стандарт
изглежда недостатъчно обмислен, а ако е вярно второто "---
непоследователен.

Замятана на руската буква „э“ с буквата „ударено и“ (\`и) изглежда
естествена.  По такъв начин се запазва буквеният характер на този
клавиш, който ще има следната история
{\usefont{X2}{cmr}{m}{n}\cyrbyus}\ar{}э\ar{}\`и.  Според БДС 5237:2006
буквата „ударено и“ се поставя на мястото на главната буква „ер
малък“~(Ь).  И след тази замяна главната буква „ер малък“ ще може да
се въвежда при активен модификатор CapsLock.

\subsubsection*{Клавишът <LSGT> (два знака)}

Далеч не всички клавиатури притежават този клавиш.  Той се намира
между клавишите ляв Shift и клавишът „Z“.  В зависимост от
клавиатурата върху него са изобразени знаците „по-голямо“ и „по-малко“
или „обратно наклонена черта“ и „вертикална черта“.

Според БДС 5237:2006 на този клавиш се поставя буквата „и с ударение“
(малка и главна).

\subsection{Фонетична клавиатура}\label{sec:phon12}

Традиционният фонетичен вариант за разположение на знаците върху
клавишите представлява изменение на американския клавиатурен стандарт
QWERTY, при който латинските букви и някои знаци са заменени с
кирилски букви.  Тъй като всички останали писмени знаци са оставени
без изменение, при традиционната фонетична клавиатура има клавиши,
които генерират малко полезните в режим „кирилица“ знаци като @, \#,
\Hat, \&, *, \_, < и~>.  Разбира се, всяка компютърна клавиатура
трябва да поддържа тези знаци, но те се използват предимно докато
клавиатурата е в режим „латиница“ и са малко полезни в режим
„кирилица“\footnote{Например ако се случи така, че в бъдеще станат
  популярни хостове в Интернет, чиито имена са на кирилица, а
  сървърният им софтуер поддържа кирилски потребителски имена, тогава
  някои от адресите за електронна поща ще са изписани на кирилица и
  значи знакът~„@“ ще се окаже частично полезен и в режим
  „кирилица“.}.  Това ни дава резерв от свободни позиции на
клавиатурата, който можем да използваме за по-полезни знаци, като
например знака за номер~(№).

Според БДС 5237:2006 на мястото на знаците „<“ и „>“ се поставят
българските кавички („\dots“).  Тази замяна е естествена, тъй като
измежду свободните знаци на традиционната фонетична клавиатура
единствено знаците „<“ и „>“ подсказват двустранност.

При разполагането върху клавишите на останалите нови знаци е разумно
да се стремим да има максимална прилика между знака, който ще се
генерира в режим „латиница“, и знака който ще се генерира в режим
„кирилица“.

Съгласно този принцип е естествена замяната на знака за номер (\#) с
българския знак за номер (№) и на знака за подчертаване (\_) с дългото
тире.  Знакът за параграф (§) пък показва известно графично сходство
със знака за амперсанд (\&).  Точно такива замени са приети и при БДС
5237:2006.

Така остават три знака, които могат да бъдат заменени (@ \Hat{} *) и
три знака, за които трябва да изберем място на клавиатурата (\euro{}
\`и \`И).  Според БДС 5237:2006 знакът за евро (\euro) се поставя на
мястото на \Hat{}, а буквата „ударено~и“ (\`{и}) "--- подобно на
клавиатурата клавиатурата тип „БДС“ "--- се поставя на мястото на
главната буква „ер малък“~(Ь).  И тук за въвеждането на главната буква
„ер малък“ трябва да се използва режимът CapsLock.

Бързо сравнение на наборите знаци, които се поддържат от двете
компютърни клавиатурни подредби в БДС 5237:2006, показва, че за сметка
на руските букви при фонетичната клавиатура се поддържат в повече
следните знаци: машинописен апостроф (${}^\prime{}$), „@“ и „*“.

На клавиша <LSGT>, който не присъства на всяка клавиатура, според БДС
5237:2006 се поставя буквата „и с ударение“ (малка и главна).

Разположението на буквените знаци при фонетичната клавиатура в БДС
5237:2006 се различава значително от обичайното.  Това прави засега
несигурно дали тази нова фонетична клавиатура ще се наложи.  Мотивите
за направените изменения са следните:

\begin{enumerate}
\item Буквата „В“ да се постави на много по-естественото място „V“
  вместо „W“, както е при традиционната фонетична клавиатура.
\item Кирилските букви без естествено фонетично съответствие в
  латинската азбука по възможност да се разположат според визуалната
  си прилика с латински букви.
\item Да се вземе предвид честотното разпределение на употребата на
  буквите в българския език, като по-често използваните букви се
  поставят на по-удобни клавиши.
\end{enumerate}

В резултат буквите при традиционната фонетична клавиатура се променят,
както следва:

\begin{itemize}
  \item Вместо „Ж“ при „V“ се поставя „В“ поради фонетично съответстие.
  \item Вместо „В“ при „W“ се поставя „Ш“ поради визуална прилика.
  \item Вместо „Ш“ при „[\{“ се поставя „Я“.
  \item Вместо „Я“ при „Q“ се поставя „Ч“ "--- малката латинска буква
    „q“ прилича на буквата „ч“.
  \item Вместо „Ч“ при „\`{~}$\tilde{}$~“ се поставя „Ю“ "--- рядко
    използвана буква и сравнително неудобен клавиш.
  \item Вместо „Ю“ при „$\backslash$|“ се поставя „Ь“ "--- най-рядко
    използваната буква се поставя на най-неудобния клавиш.
  \item Вместо „Ь“ при „X“ се поставя „Ж“ поради визуална прилика.
\end{itemize}

В този документ са дефинирани две фонетични клавиатурни подредби "---
една с традицинонното разположение на буквите и една с разположение
според БДС 5237:2006.  Небуквените знаци при фонетичната клавиатура с
традиционно разположение на буквите са разположени по същия начин,
както и при БДС 5237:2006.

\section{Знаци от трети и четвърти регистър}

\subsection{Алгоритъм за нареждане на знаците}\label{sec:algorithm}

Тъй като знаците от трети и четвърти регистър не са надписани върху
клавишите и се използват сравнително по-рядко от другите знаци, то те
трябва да се разполагат по възможно най-интуитивния и лесен за
запомняне начин.  За всеки възможен знак е търсен най-близкият до него
знак от първи или втори регистър.  За удобство ако даден знак трябва
да бъде поставн в четвърти регистър, но в трети регистър не е
дефиниран знак, то същият знак се се дублира и в трети регистър.

Знаци като по-малко или равно ($\leqq$) и градус (°), които по
принцип трябва да се въвеждат в режим „латиница“, не се поддържат в
режим „кирилица“.

Тъй като в някои случаи се оказва, че два или повече различни писмени
знака трябва да се разположат на едно и също място, то точната
клавиатурна подредба се определя посредством алгоритъм, който решава
конфликтите от този тип.  Да означим с $a_i$ знакът, който се генерира
от клавиша $a$ в $i$-ти регистър.  Тогава точното разположение на
писмените знаци от трети и четвърти регистър ще се определи от следния
алгоритъм:

\begin{enumerate}

  \item Избираме писмен знак $X$, който все още не е разположен на
    клавиатурата.

  \item Определяме писмен знак $Y$, който в известен смисъл е близък
    до $X$.  Ако забележим, че алгоритъмът „се зацикли“ и престане да
    намира места на все още неразположените писмени знаци, то
    променяме избора си на знак $Y$.

  \item Ако $Y$ не присъства на клавиатурата, преминаваме към точка~1
    и избираме друг знак $X$.  В противен случай нека $Y=a_i$.

  \item Ако $Y=a_1$ и $a_3$ не е дефиниран или е разположен някъде в
    първи или втори регистър, то $a_3:=X$ и преминаваме към точка~1.

  \item Ако $Y=a_2$ и $a_4$ не е дефиниран или е разположен някъде в
    първи или втори регистър, то $a_4:=X$ и преминаваме към точка~1.

  \item Ако $a_4$ не е дефиниран или е разположен някъде в първи или
    втори регистър, то $a_4:=X$ и преминаваме към точка~1.

  \item Ако $Y=a_4$, а $a_3$ не е дефиниран или е разположен някъде в
    първи или втори регистър, то $a_3:=a_4$, $a_4:=X$ и преминаваме
    към точка~1.

  \item Ако $a_3$ не е дефиниран или е разположен някъде в първи или
    втори регистър, то $a_3:=X$ и преминаваме към точка~1.

  \item Ако все още има знаци без място на клавиатурата, преминаваме
    към точка~1.

  \item За всеки дефиниран знак $a_4$, ако $a_3$ не е дефиниран или е
    разположен някъде в първи или втори регистър, то $a_3:=a_4$.

  \item За всеки дефиниран знак $a_3$, ако $a_4$ не е дефиниран, то
    $a_4:=a_3$.

\end{enumerate}

В следващите два подраздела са дефинирани „апроксимиращите“ знаци за
втора точка на този алгоритъм.  В първия от тях са дефинирани
апроксимиращите знаци в режим „кирилица“ (без значение коя от трите
кирилски клавиатурни подредби), а във втория "--- апроксимиращите
знаци за режим „латиница“.

\subsection{Знаци в режим „кирилица“}\label{sec:cyrsign34}

\begin{substable}
  \subsymbol{свързващо тире}{късо тире}{визуално сходство}
\end{substable}

Свързващото тире се използва тогава, когато не е желателно думата да
бъде пренесена на следващия ред.  Например при израза „д-р Петров“ не
бихме желали в края на реда да стои „д-“, а в началото на следващия
"--- „р Петров“.

\begin{substable}
  \subsymbol{средно тире (en dash)}{знак за подчертаване (\_{})}{визуално
    сходство}
  \subsymbol{дълго тире (em dash)}{=}{визуално сходство}
\end{substable}

Дългото тире ("---) се използва като синтактичен препинателен знак в
изреченията и в началото на пряка реч.  Освен това според българския
правопис дълго тире се използва и в изрази като „Ще те чакам 5---10
минути“ и „закон на Бойл---Мариот“.  През последните 20~години обаче
средното тире замени почти напълно употребата на дългото тире.  Затова
според БДС 5237:2006 между знаците от първи и втори регистър е
предвидено място за средното тире (en dash), а не за дългото тире (em
dash).

Тъй като средното тире ще има място във втори регистър, предложената
тук за него позиция от трети или четвърти регистър е по-неудобна,
отколкото позицията предвидена за дългото тире.  Все пак такава
позиция е предвидена, за да може алгоритъмът да се прилага и за
клавиатурни подредби, които не предвиждат средно тире в първи или
втори регистър, например.  традиционната фонетична клавиатура.

\begin{substable}
  \subsymbol{леви двойни кавички („)}{<}{двустранност}
  \subsymbol{десни двойни кавички (“)}{>}{двустранност}
  \subsymbol{леви френски кавички («)}{десни кавички („)}{функционалност}
  \subsymbol{десни френски кавички (»)}{леви кавички (“)}{функционалност}
\end{substable}

Клавиатурата „БДС“ не поддържа знаците за по-малко и по-голямо, но
както бе показано в раздел~\ref{sec:bds12}, при нея имаме свободен
клавиш, който може да поддържа кавичките в първи и втори регистър
(клавишът „<BKSL>“).  Фонетичната клавиатура, описана в този
документ, също поддържа кавичките на първо и второ ниво.  Тук се
предвиждат позиции от трето или четвърто ниво за обикновените двойни
кавички, за да може алгоритъмът да се прилага и при традиционната
фонетична клавиатура, която не поддържа кавичките на първи и втори
регистър.

В българския език френските кавички се използват много по-рядко.

\begin{substable}
  \subsymbol{горна запетая (’)}{маш. апостроф/запетая}{визуално сходство}
  \subsymbol{дясна единична кавичка (‘)}{горна запетая}{обща употреба}
\end{substable}

При фонетичната клавиатура поставяме тези знаци при машинописния
апостроф, а при клавиатурата тип „БДС“ "--- при запетаята (тъй като
тук няма апостроф).

Лявата единична кавичка и горната запетая са един и същ символ,
известен също и като апостроф.  Обърнете внимание, че това не е
знакът, известен като машинописен апостроф (${}^\prime{}$) и който в режим
„латиница“ излиза при натискането на предпоследния клавиш от третия
ред на клавиатурата и прилича на къса вертикална чертица.  Горната
запетая (апострофът, дясната единична кавичка), изглежда като запетая,
повдигната наравно с височината на текста (например „Олимпиада’2000“).

На български език единичните кавички се използват предимно в научни
текстове при пояснение на значението на дума или израз.  Според
последното издание на правописния речник българските единични кавички
са ‘\dots’.  Това обаче е спорно, тъй като почти всички книги издадени
до момента научни книги, включително и при предходната редакция на
правописния речник, единичните кавички са ’\dots’, т.е. отварящата и
затварящата са еднакви и изглеждат просто като повдигнати запетаи.
Във връзка с това възниква въпросът дали в новия вариант на
правописния речник промяната е направена съзнателно и обмислено или
пък просто е грешка, допусната при предпечатната подготовка, с която
авторите на речника нямат нищо общо.

\begin{substable}
  \subsymbol{многоточие (\dots)}{двоеточие (:)}{визуално сходство}
\end{substable}

Разликата между многоточието (\dots) и трите последователни точки
(...) е това, че при многоточието разтоянието между отделните точки е
по-голямо.  При използването на непропорционален шрифт обаче
(напр. „Courier“, „FreeMono“, „DejaVu Sans Mono“ и „Nimbus Mono“)
многоточието ще бъде ненормално сбито и за това при тавива шрифтове е
по-добре да се използват три последователни точки вместо многочие.

\begin{substable}
  \subsymbol{Лява скоба за коментари ([)}{(}{визуално сходсво}
    \subsymbol{Дясна скоба за коментари (])}{)}{визуално сходсво}
  \subsymbol{Лява скоба за коментари ($\langle$)}{[ (фонет.) или \% (БДС)}{}
    \subsymbol{Дясна скоба за коментари ($\rangle$)}{] (фонет.) или
    $\langle$ (БДС)}{функционалност}
\end{substable}

Когато цитираме нечий текст и искаме да вмъкнем в него собствен
коментар, но така че да е ясно, че коментарът не е на оригиналният
автор, се използват квадратни или ъглови скоби.  С [\dots] или
$\langle\dots\rangle$ пък посочваме, че сме изпуснали част от
авторовия текст.  Използването на квадратните скоби е за предпочитане,
тъй като много от шрифтовете не поддържат ъгловите скоби.  Затова
предложените тук апроксимационни замени предвиждат по-удобно място за
квадратните скоби.

При фонетичната клавиатура поставяме ъгловите скоби при квадратните.
Това е възможно, защото тази клавиатура използва различни клавиши за
отварящата и за затварящата скоба.  По такъв начин можем да поставим
квадратните скоби в трети регистър, а ъгловите "--- в четвърти.  При
клавиатурата тип „БДС“ обаче квадратните скоби са поставени на един
клавиш.  Затова тук за ъгловите скоби се използва клавишът със знак за
процент.

\begin{substable}
  \subsymbol{свързващ интервал}{интервал}{визуална идентичност}
\end{substable}

Свързващият интервал се използва, когато не желаем изразът да бъде
пренесен на следващия ред.  Например при израза „с.~Яхиново“ няма да
ни се иска в края на реда да стои „с.“, а в началото на следващия "---
„Яхиново“.  Затова след „с.“ използваме свързващ интервал.

Визуално свързващият интервал представлява най-обикновен интервал.  За
съжаление както при OpenOffice.org, така и при Abiword (а също и при
най-известния несвободен офиспакет) свързващият интервал е винаги с
една и съща ширина.  Т.е. при режим „двустранно подравняване“
обикновеният интервал ще се разтегли, но не и свързващият.  Според
специалисти по предпечат това е неправилно.

\begin{substable}
  \subsymbol{№}{\#}{знак за номер}
  \subsymbol{§}{\&}{визуална прилика}
  \subsymbol{\euro{}}{\$}{парична единица}
\end{substable}

При клавиатурата тип „БДС“, както и при тук описаните фонетични
клавиатури, тези три знака се поддържат в първи или втори регистър.
Тук за тях са предвидени „апроксимиращи“ знаци, за да може алгоритъмът
да се прилага и за клавиатури като традиционната фонетична клавиатура,
при които тези знаци не се поддържат.

Българският знак за номер се поставя при знакът за диез, който в някои
езици също изпълнява функциите на знак за номер, а знакът за параграф
"--- при знака за амперсанд, който показва известно визуално сходство.
Знакът за евро се поставя за знака за долар, който също е парична
единица.

\begin{substable}
  \subsymbol{знак за починали хора ($\dagger$)}{+}{визуална прилика}
\end{substable}

Знакът за починали хора ($\dagger$) се поставя при датата на смъртта,
а понякога при името на починали хора.  При някои шрифтове изглежда
като кръст, но по принцип се различава от знака за кръст.

\begin{substable}
  \subsymbol{знак за авторски права ({\usefont{T1}{cmr}{m}{n}\copyright})}{с}{визуално сходство}
  \subsymbol{знак за авторски права ({\usefont{T1}{cmr}{m}{n}\copyright})}{к (БДС) или ц
    (фонет.)}{фонетично сходство}
\end{substable}

За по-лесно намиране, знакът за авторски права се поставя на два
различни клавиша.  Първо, той се поставя при кирилската буква „с“, тъй
като тя прилича на буквичката в кръгчето.  Второ, при фонетичната
клавиатура той се поставя при „ц“, тъй като там е и латинската буква
„c“.  При клавиатурата „БДС“ е неестествено да обръщаме внимание на
латинските букви и затова знакът там е дублиран при буквата „к“ (от
транскрибирания вариант на copyright "--- копирайт).

\begin{substable}
  \subsymbol{регистр. търг. марка (®)}{р}{фонетично сходство}
  \subsymbol{търговска марка ({\small {™}})}{т}{фонетично сходство}
\end{substable}

Знакът ™ няма нито законови основания, нито пък някаква сериозна
традиция в България.  Въпреки това е възможно някои фирми да пожелаят
да го използват за свои нерегистрирани търговски марки и затова за
този знак също е осигурена поддръжка.

\begin{substable}
  \subsymbol{малко и с ударение (\`{и})}{и}{варианти}
  \subsymbol{главно и с ударение (\`{И})}{\`{и}}{малка/главна буква}
\end{substable}

Тъй като и двете клавиатури, дефинирани в този документ, поддържат
буквата „и с надреден знак във вид на ударение“ във втори регистър,
поставянето му в трети регистър има смисъл единствено при прилагането
на този алгоритъм към традиционната фонетична клавиатура.

\begin{substable}
  \subsymbol{ы}{ь}{визуално сходство}
  \subsymbol{Ы}{ы}{малка/главна буква}
  \subsymbol{э}{e}{звуково сходство}
  \subsymbol{Э}{э}{малка/главна буква}
\end{substable}

При клавиатурата „БДС“ руските букви са поставени на първо и второ
ниво, така че за тях е предвидено място на трето и четвърто ниво
заради фонетичната клавиатура.

\begin{substable}
  \subsymbol{{\usefont{X2}{cmr}{m}{n}\cyryat}}{я}{името на буквата}

  \subsymbol{{\usefont{X2}{cmr}{m}{n}\CYRYAT}}
            {{\usefont{X2}{cmr}{m}{n}\cyryat}}{малка/главна буква}

  \subsymbol{{\usefont{X2}{cmr}{m}{n}\cyrbyus}}{ъ}{звукова идентичност}

  \subsymbol{{\usefont{X2}{cmr}{m}{n}\CYRBYUS}}
            {{\usefont{X2}{cmr}{m}{n}\cyrbyus}}{малка/главна буква}

  \subsymbol{{\scriptsize{I}}{\usefont{X2}{cmr}{m}{n}\cyrbyus}}{й}{звуково
    подобие}

  \subsymbol{I{\usefont{X2}{cmr}{m}{n}\CYRBYUS}}
            {{\scriptsize{I}}{\usefont{X2}{cmr}{m}{n}\cyrbyus}}
            {малка/главна буква}
\end{substable}

Това са букви, които са се използвали преди правописната реформа през
1945~г.  Буквата {\usefont{X2}{cmr}{m}{n}\cyryat} е старата буква
„ят“, която често се нарича „променливо я“ и затова отива при „я“.
Големият юс ({\usefont{X2}{cmr}{m}{n}\cyrbyus}) в книжовния български
език винаги се чете „ъ“ и затова отива при „ъ“.  Йотираният голям юс
({\scriptsize{I}}{\usefont{X2}{cmr}{m}{n}\cyrbyus}) се използвал
при сравнително малко от правописните варианти преди 1945~г.  Той
отива при „й“, защото и двете букви са йотирани.

\begin{substable}
  \subsymbol{остро ударение (\`{})}{/}{визуална прилика}
  \subsymbol{тежко ударение (\'{})}{остро ударение}{функционално идентични}
\end{substable}

Тези знаци се използват за поставяне на ударение над произволна буква
и не се поддържат от всички редактори.  При знака „/“ се поставя
острото, а не тежкото ударение, за да бъде позицията на острото
ударение по-достъпна.  Българската полиграфска традиция почти винаги
използва остро, а не тежко ударение, като смисълът и на двете е
абсолютно идентичен (при някои други езици има разлика между двете
ударения).

\subsection{Знаци в режим „латиница“}\label{sec:latsign34}

В режим „латиница“ се поддържат препинателните знаци, необходими при
въвеждането на английски текст, както и някои технически математически
знаци.

При предложените тук апроксимационни замени се получава разширена
клавиатурна подредба за писане на английски език.  Не е предвидена
възможност за въвеждане на разнообразните неанглийски латински букви,
тъй като едва ли е възможно с една единствена латинска клавиатурна
подредба да се обхванат по удобен начин всички латински букви от
по-разпространените европейски езици.

\subsubsection{Препинателни знаци}

\begin{substable}

  \subsymbol{свързващо тире}{късо тире}{визуално сходство}

  \subsymbol{дълго тире (em dash)}{=}{визуално сходство}

  \subsymbol{средно тире (en dash)}{знак за подчертаване}{визуално сходство}

\end{substable}

Това че в режим „латиница“ средното тире не е измежду знаците от първи
или втори регистър, a е разположено на по-неудобно място в сравнение с
дългото тире (в четвърти вместо трети регистър) не е проблем, защото
според повечето американски източници, както и според
по-консервативните британски, в почти всички случаи трябва да се
използва дълго, а не средно тире, като при това за разлика от
българската традиция, то не е отделено с интервали.  Следният цитат от
книгата на Leslie Lamport „\LaTeX : a document preparation system“
илюстрира употребата на тиретата в английския език:
\begin{quote}
  An intra-word dash or hyphen, as in X-ray. \\
  A medium dash as for number ranges, like 1--2.\\
  A punctuation dash---like this.
\end{quote}

\begin{substable}

  \subsymbol{дясна единич. кавичка (`)}{/}{визуално сходство}

  \subsymbol{лява единич. кавичка (')}{дясна единич. кавичка (`)}{обща употреба}

  \subsymbol{десни двойни кавички (``)}{обикн. двойни кавички}{обща функционалност}

  \subsymbol{леви двойни кавички ('')}{десни кавички (``)}{свързана употреба}

  \subsymbol{$\bullet$}{@}{визуално сходство}

  \subsymbol{\S}{\&}{визуално сходство}


\end{substable}

Има разлика между двойните кавички в режим „латиница“ (``\dots'') и
кавичките в режим кирилица („\dots“).  Кръгчето (bullet, $\bullet$) не
се поддържа в режим кирилица, защото българската правописна традиция
изисква използването на тирета за неномерирани списъци.  От друга
страна знакът за номер~(№) не се поддържа в режим латиница, тъй като е
непознат в англоезичната писмена традиция.

\begin{substable}

  \subsymbol{\copyright}{c}{визуално сходство}

  \subsymbol{®}{r}{визуално сходство}

  \subsymbol{\small {™}}{t}{визуално сходство}

\end{substable}

Знаците за авторски права, регистрирана търговска марка и
нерегистрирана търговска марка естествено си намират място според
латинската буква, която съдържат.

\begin{substable}

  \subsymbol{\euro}{\$}{парична единица}

  \subsymbol{{\usefont{T1}{cmr}{m}{n}\pounds}}{\euro}{парична единица}

  \subsymbol{\yen}{y}{визуално сходство}

\end{substable}

Поддържат се знаците на паричните единици щатски долар, евро,
британска лира и йена.

\subsubsection{Математически знаци}

\samepage{
\begin{substable}
  \subsymbol{градус (\textdegree)}{o}{визуално сходство}

  \subsymbol{промил ({\usefont{T1}{cmr}{m}{n}\textperthousand})}{\%}{визуално сходство}

  \subsymbol{втора степен (${}^2$)}{2}{визуално сходство}

  \subsymbol{трета степен (${}^3$)}{3}{визуално сходство}

  \subsymbol{n-та степен (${}^n$)}{n}{визуално сходство}

  \subsymbol{$\pm$}{+}{визуално сходство}

  \subsymbol{$\times$}{*}{близка функционалност}

  \subsymbol{$\div$}{:}{близка функционалност}

\end{substable}
}

Тези знаци често се срещат и при нематематически и нетехнически
текстове.  Когато знакът за градус се използва като мярка за големина
на ъгъл, между него и предходното число не се слага интервал (например
ъгъл от 45\textdegree).  Ако обаче знакът за градус се използва за
температура, тогава между числото и знака за градус се слага малко
интервалче (например температура от 20\,\textdegree{}C).

Знаците за степени трябва да се използват единствено при писане на
прости неформатирани текстове, например при електронната поща.  В
останалите случаи е по-добре да се използват възможностите на
използваната програма за редактиране на електронни документи.

\begin{substable}

  \subsymbol{$\Vert$}{h}{визуално сходство}

  \subsymbol{$\perp$}{|}{перпендикуляр}

  \subsymbol{$\angle$}{z}{визуално сходство}

  \subsymbol{$\infty$}{8}{визуално сходство}

  \subsymbol{$\int$}{i}{името на знака}

  \subsymbol{$\oint$}{O}{визуално сходство}

  \subsymbol{$\partial$}{9}{визуално сходство}

  \subsymbol{$\nabla$}{$\partial$}{свързана употреба}

\end{substable}

Знакът за обикновен интеграл по естествен начин отива при буквата „i“.
Знакът за кръгов интеграл отива при главната буква „O“, защото при
малката буква е знакът за градус.  От всички латински букви,
най-голяма прилика до „$\Vert$“ и „$\angle$“ показват съответно
буквите „H“ и „Z“.  Цифрите 8 и 9 приличат на знаците за безкрайност
($\infty$) и частна производна ($\partial$).

\begin{substable}

  \subsymbol{$\neq$}{\#}{визуално сходство}

  \subsymbol{$\leqq$}{<}{визуално сходство}

  \subsymbol{$\geqq$}{>}{визуално сходство}

  \subsymbol{$\approx$}{$\tilde{}$}{визуално сходство}

  \subsymbol{$\cong$}{$\approx$}{визуално сходство}

  \subsymbol{$\equiv$}{$\backslash$}{}

\end{substable}

Знакът за всяка от тези математически релации по естествен начин отива
при най-близкия до него знак.  Изключение е знакът за еквивалентност
($\equiv$), който не може да отиде при знака за равенство, тъй като
там е значително по-често използваният знак дълго тире.

Използването на знаците „$\leqq$“ и „$\geqq$“ вместо
по-разпространените „$\leqslant$“ и „$\geqslant$“ е продиктувано от
следните съображения:

\begin{enumerate}
\item При много от разпространените в момента компютърни шрифтове
  знаците „$\leqslant$“ и „$\geqslant$“ не са дефинирани.
\item Преди двадесетина години научното математическо списание
  „Pliska“, издавано от БАН, задължаваше авторите да използват
  „$\leqq$“ и „$\geqq$“, а не някой от другите варианти за знаците за
  по-малко и по-голямо.  Това означава, че употребата на „$\leqq$“ и
  „$\geqq$“ не само е допустима, но според някои математици от БАН е и
  единствено правилна.
\end{enumerate}

\begin{substable}

  \subsymbol{$\lnot$}{!}{близка функционалност}

  \subsymbol{$\wedge$}{$\hat{}$}{визуално сходство}

  \subsymbol{$\vee$}{$\wedge{}$}{визуално сходство}

  \subsymbol{$\rightarrow$}{$\geqq$}{визуално сходство}

  \subsymbol{$\leftrightarrow$}{$\leqq$}{визуално сходство}

  \subsymbol{$\forall$}{a}{визуално сходство}

  \subsymbol{$\exists$}{e}{визуално сходство}

\end{substable}

Логическите знаци също могат да се разположат по естествен начин, като
изключение донякъде е само знакът за отрицание ($\lnot$).  Той отива
при знака за унивителна, тъй като при повечето от най-популярните в
момента езици за програмиране (напр. Си, Джава, Пърл) отрицанието се
обозначава с удивителна.

\begin{substable}

  \subsymbol{$\in$}{]}{}

  \subsymbol{$\not\in$}{$\in$}{визуално сходство}

  \subsymbol{$\subseteq$}{[}{визуално сходство}

  \subsymbol{$\subset$}{$\subseteq$}{визуално сходство}

  \subsymbol{$\varnothing$}{0}{близко значение}

  \subsymbol{$\cup$}{U}{визуално сходство}

  \subsymbol{$\cap$}{u}{визуално сходство}

\end{substable}

Математическите знаци за множества също се поддържат.

\begin{substable}
  \subsymbol{$\alpha, \beta, \gamma, \delta, \varepsilon, \zeta, \iota,
    \kappa, \lambda$}{A, B, G, D, E, Z, I, K, L}{името на буквата}

  \subsymbol{$\mu, \nu, \xi, \pi, \rho, \sigma, \tau,
    \varphi$}{M, N, X, P, R, S, T, F}{името на буквата}

  \subsymbol{$\Gamma, \Delta, \Lambda, \Xi, \Pi, \Sigma, \Phi$}{g, d, l, x,
    p, s, f}{името на буквата}

  \subsymbol{$\eta, \upsilon$}{H, Y}{вида на главната буква}

  \subsymbol{$\omega, \Omega$}{W, w}{вида на малката буква}

  \subsymbol{$\theta, \Theta, \psi, \Psi$}{Q, q, V, v}{прилика}

  \subsymbol{$\chi$}{C}{името на английски (chi)}

  \subsymbol{$\varkappa$}{k}{името на буквата}

  \subsymbol{$\vartheta$}{b}{}

\end{substable}

Много от гръцките букви редовно се използват в технически текстове.
Затова и за тях е предвидена възможност те да се въвеждат директно от
клавиатурата.  Някои от гръцките букви почти никога не се използват в
технически текстове, за такива букви не са предвидени позиции на
клавиатурата.  Това дава възможност освободените клавишни позиции да
се използват за изброените вече разнообразни други знаци.

Главните гръцки букви са поставени в трети регистър, а малките "--- в
четвърти.  Макар и малко неестествено, това е направено, защото
повечето от главните гръцки букви не се използват в технически
текстове.  По такъв начин си осигуряваме повече свободни позиции от
трети регистър и останалите знаци могат да се разположат на по-удобни
места.

\section{Формална дефиниция на разположението  на знаците върху клавишите}

Таблиците от този раздел дават формална дефиниция на разположението на
знаците върху клавишите на клавиатурата.  По-нагледни изображения има
в приложение~\ref{sec:ilustrations}.

Използваните имена на клавишите сa въз основа на ISO 9995 и са
илюстрирани в следното изображение:

\begin{keyboard}

\hline

\keys{E00}& \keys{E01}& \keys{E02}& \keys{E03}& \keys{E04}& \keys{E05}&
\keys{E06}& \keys{E07}& \keys{E08}& \keys{E09}& \keys{E10}& \keys{E11}&
\keys{E12}& \keyspace{5}{$\leftarrow$} \\

\keyn{\`{~}}& \keyn{1}& \keyn{2}& \keyn{3}& \keyn{4}& \keyn{5}& \keyn{6}&
\keyn{7}& \keyn{8}& \keyn{9}& \keyn{0}& \keyn{-}& \keyn{=}& 
\keyspace{5}{\keyname{Backspace}} \\

\hline

\keyspace{4}{$\leftrightarrows$}& \keys{D01}& \keys{D02}& \keys{D03}& \keys{D04}&
\keys{D05}& \keys{D06}& \keys{D07}& \keys{D08}& \keys{D09}& \keys{D10}& \keys{D11}&
\keys{D12}& \keyspace{4}{\keyname{Enter}} \\

\keyspace{4}{\keyname{Tab}}& \keyn{}& \keyn{}& \keyn{}& \keyn{}&
\keyn{}& \keyn{}& \keyn{}& \keyn{}& \keyn{}& \keyn{}& \keyn{[}&
\keyn{]}& \keyspace{4}{$\hookleftarrow$} \\

\cline{1-41}

\keyspace{5}{$\Uparrow$\keyname{Caps}}& \keys{C01}& \keys{C02}& \keys{C03}& \keys{CO4}&
\keys{C05}& \keys{C06}& \keys{C07}& \keys{C08}& \keys{C09}& \keys{C10}& \keys{C11}&
\keys{C12}& \keyspace{3}{} \\

\keyspace{5}{\keyname{Lock}}& \keyn{}& \keyn{}& \keyn{}&
\keyn{}& \keyn{}& \keyn{}& \keyn{}& \keyn{}& \keyn{}& \keyn{;}&
\keyn{${}^\prime$}& \keyn{$\backslash$}& \keyspace{3}{} \\

\hline

\keyspace{4}{$\uparrow$}& \keys{B00}& \keys{B01}& \keys{B02}& \keys{B03}& \keys{B04}&
\keys{B05}& \keys{B06}& \keys{B07}& \keys{B08}& \keys{B09}& \keys{B10}&
\keyspace{7}{$\uparrow$} \\

\keyspace{4}{\keyname{Shift}}& \keyn{<}& \keyn{}& \keyn{}& \keyn{}& \keyn{}&
\keyn{}& \keyn{}& \keyn{}& \keyn{,}& \keyn{.}&
\keyn{/}& \keyspace{7}{\keyname{Shift}} \\

\hline

\end{keyboard}

\subsection{Вариант тип „БДС“ за разположението на знаците}

\begin{longtable}{lllll}
  \toprule
   & \multicolumn{2}{c}{\tabtit{Unicode}}
   & \multicolumn{2}{c}{\tabtit{X keysym}} \\
  \cmidrule(rl){2-3}
  \cmidrule(l){4-5}
  \tabtit{Клавиш} & \tabtit{1 и 2 рег.} & \tabtit{3 и 4 рег.}
                  & \tabtit{1 и 2 рег.} & \tabtit{3 и 4 рег.} \\
  \midrule
  \endhead
  \bottomrule
  \endfoot
E00 & U+0028 & U+005B & \scriptsize{parenleft} & \scriptsize{bracketleft} \\
     & U+0029 & U+005D & \scriptsize{parenright} & \scriptsize{bracketright} \\
E01 & U+0031 &  & \scriptsize{1} & \scriptsize{} \\
     & U+0021 &  & \scriptsize{exclam} & \scriptsize{} \\
E02 & U+0032 &  & \scriptsize{2} & \scriptsize{} \\
     & U+003F &  & \scriptsize{question} & \scriptsize{} \\
E03 & U+0033 & U+2020 & \scriptsize{3} & \scriptsize{dagger} \\
     & U+002B & U+2020 & \scriptsize{plus} & \scriptsize{dagger} \\
E04 & U+0034 &  & \scriptsize{4} & \scriptsize{} \\
     & U+0022 &  & \scriptsize{quotedbl} & \scriptsize{} \\
E05 & U+0035 & U+2329 & \scriptsize{5} & \scriptsize{U2329} \\
     & U+0025 & U+232A & \scriptsize{percent} & \scriptsize{U232A} \\
E06 & U+0036 & U+2014 & \scriptsize{6} & \scriptsize{emdash} \\
     & U+003D & U+2014 & \scriptsize{equal} & \scriptsize{emdash} \\
E07 & U+0037 & U+2026 & \scriptsize{7} & \scriptsize{ellipsis} \\
     & U+003A & U+2026 & \scriptsize{colon} & \scriptsize{ellipsis} \\
E08 & U+0038 & U+0300 & \scriptsize{8} & \scriptsize{U0300} \\
     & U+002F & U+0301 & \scriptsize{slash} & \scriptsize{U0301} \\
E09 & U+0039 &  & \scriptsize{9} & \scriptsize{} \\
     & U+2013 &  & \scriptsize{endash} & \scriptsize{} \\
E10 & U+0030 &  & \scriptsize{0} & \scriptsize{} \\
     & U+2116 &  & \scriptsize{numerosign} & \scriptsize{} \\
E11 & U+002D & U+2011 & \scriptsize{minus} & \scriptsize{U2011} \\
     & U+0024 & U+20AC & \scriptsize{dollar} & \scriptsize{EuroSign} \\
E12 & U+002E &  & \scriptsize{period} & \scriptsize{} \\
     & U+20AC &  & \scriptsize{EuroSign} & \scriptsize{} \\
D01 & U+002C & U+2019 & \scriptsize{comma} & \scriptsize{rightsinglequotemark} \\
     & U+044B & U+2018 & \scriptsize{Cyrillic\_yeru} & \scriptsize{leftsinglequotemark} \\
D02 & U+0443 &  & \scriptsize{Cyrillic\_u} & \scriptsize{} \\
     & U+0423 &  & \scriptsize{Cyrillic\_U} & \scriptsize{} \\
D03 & U+0435 & U+044D & \scriptsize{Cyrillic\_ie} & \scriptsize{Cyrillic\_e} \\
     & U+0415 & U+042D & \scriptsize{Cyrillic\_IE} & \scriptsize{Cyrillic\_E} \\
D04 & U+0438 & U+045D & \scriptsize{Cyrillic\_i} & \scriptsize{U045D} \\
     & U+0418 & U+040D & \scriptsize{Cyrillic\_I} & \scriptsize{U040D} \\
D05 & U+0448 &  & \scriptsize{Cyrillic\_sha} & \scriptsize{} \\
     & U+0428 &  & \scriptsize{Cyrillic\_SHA} & \scriptsize{} \\
D06 & U+0449 &  & \scriptsize{Cyrillic\_shcha} & \scriptsize{} \\
     & U+0429 &  & \scriptsize{Cyrillic\_SHCHA} & \scriptsize{} \\
D07 & U+043A & U+00A9 & \scriptsize{Cyrillic\_ka} & \scriptsize{copyright} \\
     & U+041A & U+00A9 & \scriptsize{Cyrillic\_KA} & \scriptsize{copyright} \\
D08 & U+0441 & U+00A9 & \scriptsize{Cyrillic\_es} & \scriptsize{copyright} \\
     & U+0421 & U+00A9 & \scriptsize{Cyrillic\_ES} & \scriptsize{copyright} \\
D09 & U+0434 &  & \scriptsize{Cyrillic\_de} & \scriptsize{} \\
     & U+0414 &  & \scriptsize{Cyrillic\_DE} & \scriptsize{} \\
D10 & U+0437 &  & \scriptsize{Cyrillic\_ze} & \scriptsize{} \\
     & U+0417 &  & \scriptsize{Cyrillic\_ZE} & \scriptsize{} \\
D11 & U+0446 &  & \scriptsize{Cyrillic\_tse} & \scriptsize{} \\
     & U+0426 &  & \scriptsize{Cyrillic\_TSE} & \scriptsize{} \\
D12 & U+003B &  & \scriptsize{semicolon} & \scriptsize{} \\
     & U+00A7 &  & \scriptsize{section} & \scriptsize{} \\
C01 & U+044C & U+044B & \scriptsize{Cyrillic\_softsign} & \scriptsize{Cyrillic\_yeru} \\
     & U+045D & U+042B & \scriptsize{U045D} & \scriptsize{Cyrillic\_YERU} \\
C02 & U+044F & U+0463 & \scriptsize{Cyrillic\_ya} & \scriptsize{U0463} \\
     & U+042F & U+0462 & \scriptsize{Cyrillic\_YA} & \scriptsize{U0462} \\
C03 & U+0430 &  & \scriptsize{Cyrillic\_a} & \scriptsize{} \\
     & U+0410 &  & \scriptsize{Cyrillic\_A} & \scriptsize{} \\
C04 & U+043E &  & \scriptsize{Cyrillic\_o} & \scriptsize{} \\
     & U+041E &  & \scriptsize{Cyrillic\_O} & \scriptsize{} \\
C05 & U+0436 &  & \scriptsize{Cyrillic\_zhe} & \scriptsize{} \\
     & U+0416 &  & \scriptsize{Cyrillic\_ZHE} & \scriptsize{} \\
C06 & U+0433 &  & \scriptsize{Cyrillic\_ghe} & \scriptsize{} \\
     & U+0413 &  & \scriptsize{Cyrillic\_GHE} & \scriptsize{} \\
C07 & U+0442 & U+2122 & \scriptsize{Cyrillic\_te} & \scriptsize{trademark} \\
     & U+0422 & U+2122 & \scriptsize{Cyrillic\_TE} & \scriptsize{trademark} \\
C08 & U+043D &  & \scriptsize{Cyrillic\_en} & \scriptsize{} \\
     & U+041D &  & \scriptsize{Cyrillic\_EN} & \scriptsize{} \\
C09 & U+0432 &  & \scriptsize{Cyrillic\_ve} & \scriptsize{} \\
     & U+0412 &  & \scriptsize{Cyrillic\_VE} & \scriptsize{} \\
C10 & U+043C &  & \scriptsize{Cyrillic\_em} & \scriptsize{} \\
     & U+041C &  & \scriptsize{Cyrillic\_EM} & \scriptsize{} \\
C11 & U+0447 &  & \scriptsize{Cyrillic\_che} & \scriptsize{} \\
     & U+0427 &  & \scriptsize{Cyrillic\_CHE} & \scriptsize{} \\
C12 & U+201E & U+00AB & \scriptsize{doublelowquotemark} & \scriptsize{guillemotleft} \\
     & U+201C & U+00BB & \scriptsize{leftdoublequotemark} & \scriptsize{guillemotright} \\
B00 & U+045D &  & \scriptsize{U045D} & \scriptsize{} \\
     & U+040D &  & \scriptsize{U040D} & \scriptsize{} \\
B01 & U+044E &  & \scriptsize{Cyrillic\_yu} & \scriptsize{} \\
     & U+042E &  & \scriptsize{Cyrillic\_YU} & \scriptsize{} \\
B02 & U+0439 & U+046D & \scriptsize{Cyrillic\_shorti} & \scriptsize{U046D} \\
     & U+0419 & U+046C & \scriptsize{Cyrillic\_SHORTI} & \scriptsize{U046C} \\
B03 & U+044A & U+046B & \scriptsize{Cyrillic\_hardsign} & \scriptsize{U046B} \\
     & U+042A & U+046A & \scriptsize{Cyrillic\_HARDSIGN} & \scriptsize{U046A} \\
B04 & U+044D &  & \scriptsize{Cyrillic\_e} & \scriptsize{} \\
     & U+042D &  & \scriptsize{Cyrillic\_E} & \scriptsize{} \\
B05 & U+0444 &  & \scriptsize{Cyrillic\_ef} & \scriptsize{} \\
     & U+0424 &  & \scriptsize{Cyrillic\_EF} & \scriptsize{} \\
B06 & U+0445 &  & \scriptsize{Cyrillic\_ha} & \scriptsize{} \\
     & U+0425 &  & \scriptsize{Cyrillic\_HA} & \scriptsize{} \\
B07 & U+043F &  & \scriptsize{Cyrillic\_pe} & \scriptsize{} \\
     & U+041F &  & \scriptsize{Cyrillic\_PE} & \scriptsize{} \\
B08 & U+0440 & U+00AE & \scriptsize{Cyrillic\_er} & \scriptsize{registered} \\
     & U+0420 & U+00AE & \scriptsize{Cyrillic\_ER} & \scriptsize{registered} \\
B09 & U+043B &  & \scriptsize{Cyrillic\_el} & \scriptsize{} \\
     & U+041B &  & \scriptsize{Cyrillic\_EL} & \scriptsize{} \\
B10 & U+0431 &  & \scriptsize{Cyrillic\_be} & \scriptsize{} \\
     & U+0411 &  & \scriptsize{Cyrillic\_BE} & \scriptsize{} \\
интервал & U+0020 & U+00A0 & \scriptsize{space} & \scriptsize{nobreakspace} \\
     & U+0020 & U+00A0 & \scriptsize{space} & \scriptsize{nobreakspace} \\
\end{longtable}

\subsection{Фонетичен вариант за разположението на знаците}

\begin{longtable}{lllll}
  \toprule
   & \multicolumn{2}{c}{\tabtit{Unicode}}
   & \multicolumn{2}{c}{\tabtit{X keysym}} \\
  \cmidrule(rl){2-3}
  \cmidrule(l){4-5}
  \tabtit{Клавиш} & \tabtit{1 и 2 рег.} & \tabtit{3 и 4 рег.}
                  & \tabtit{1 и 2 рег.} & \tabtit{3 и 4 рег.} \\
  \midrule
  \endhead
  \bottomrule
  \endfoot
E00 & U+0447 &  & \scriptsize{Cyrillic\_che} & \scriptsize{} \\
     & U+0427 &  & \scriptsize{Cyrillic\_CHE} & \scriptsize{} \\
E01 & U+0031 &  & \scriptsize{1} & \scriptsize{} \\
     & U+0021 &  & \scriptsize{exclam} & \scriptsize{} \\
E02 & U+0032 &  & \scriptsize{2} & \scriptsize{} \\
     & U+0040 &  & \scriptsize{at} & \scriptsize{} \\
E03 & U+0033 &  & \scriptsize{3} & \scriptsize{} \\
     & U+2116 &  & \scriptsize{numerosign} & \scriptsize{} \\
E04 & U+0034 & U+20AC & \scriptsize{4} & \scriptsize{EuroSign} \\
     & U+0024 & U+20AC & \scriptsize{dollar} & \scriptsize{EuroSign} \\
E05 & U+0035 &  & \scriptsize{5} & \scriptsize{} \\
     & U+0025 &  & \scriptsize{percent} & \scriptsize{} \\
E06 & U+0036 &  & \scriptsize{6} & \scriptsize{} \\
     & U+20AC &  & \scriptsize{EuroSign} & \scriptsize{} \\
E07 & U+0037 &  & \scriptsize{7} & \scriptsize{} \\
     & U+00A7 &  & \scriptsize{section} & \scriptsize{} \\
E08 & U+0038 &  & \scriptsize{8} & \scriptsize{} \\
     & U+002A &  & \scriptsize{asterisk} & \scriptsize{} \\
E09 & U+0039 & U+005B & \scriptsize{9} & \scriptsize{bracketleft} \\
     & U+0028 & U+2329 & \scriptsize{parenleft} & \scriptsize{U2329} \\
E10 & U+0030 & U+005D & \scriptsize{0} & \scriptsize{bracketright} \\
     & U+0029 & U+232A & \scriptsize{parenright} & \scriptsize{U232A} \\
E11 & U+002D & U+2011 & \scriptsize{minus} & \scriptsize{U2011} \\
     & U+2013 & U+2011 & \scriptsize{endash} & \scriptsize{U2011} \\
E12 & U+003D & U+2014 & \scriptsize{equal} & \scriptsize{emdash} \\
     & U+002B & U+2020 & \scriptsize{plus} & \scriptsize{dagger} \\
D01 & U+044F & U+0463 & \scriptsize{Cyrillic\_ya} & \scriptsize{U0463} \\
     & U+042F & U+0462 & \scriptsize{Cyrillic\_YA} & \scriptsize{U0462} \\
D02 & U+0432 &  & \scriptsize{Cyrillic\_ve} & \scriptsize{} \\
     & U+0412 &  & \scriptsize{Cyrillic\_VE} & \scriptsize{} \\
D03 & U+0435 & U+044D & \scriptsize{Cyrillic\_ie} & \scriptsize{Cyrillic\_e} \\
     & U+0415 & U+042D & \scriptsize{Cyrillic\_IE} & \scriptsize{Cyrillic\_E} \\
D04 & U+0440 & U+00AE & \scriptsize{Cyrillic\_er} & \scriptsize{registered} \\
     & U+0420 & U+00AE & \scriptsize{Cyrillic\_ER} & \scriptsize{registered} \\
D05 & U+0442 & U+2122 & \scriptsize{Cyrillic\_te} & \scriptsize{trademark} \\
     & U+0422 & U+2122 & \scriptsize{Cyrillic\_TE} & \scriptsize{trademark} \\
D06 & U+044A & U+046B & \scriptsize{Cyrillic\_hardsign} & \scriptsize{U046B} \\
     & U+042A & U+046A & \scriptsize{Cyrillic\_HARDSIGN} & \scriptsize{U046A} \\
D07 & U+0443 &  & \scriptsize{Cyrillic\_u} & \scriptsize{} \\
     & U+0423 &  & \scriptsize{Cyrillic\_U} & \scriptsize{} \\
D08 & U+0438 & U+045D & \scriptsize{Cyrillic\_i} & \scriptsize{U045D} \\
     & U+0418 & U+040D & \scriptsize{Cyrillic\_I} & \scriptsize{U040D} \\
D09 & U+043E &  & \scriptsize{Cyrillic\_o} & \scriptsize{} \\
     & U+041E &  & \scriptsize{Cyrillic\_O} & \scriptsize{} \\
D10 & U+043F &  & \scriptsize{Cyrillic\_pe} & \scriptsize{} \\
     & U+041F &  & \scriptsize{Cyrillic\_PE} & \scriptsize{} \\
D11 & U+0448 &  & \scriptsize{Cyrillic\_sha} & \scriptsize{} \\
     & U+0428 &  & \scriptsize{Cyrillic\_SHA} & \scriptsize{} \\
D12 & U+0449 &  & \scriptsize{Cyrillic\_shcha} & \scriptsize{} \\
     & U+0429 &  & \scriptsize{Cyrillic\_SHCHA} & \scriptsize{} \\
C01 & U+0430 &  & \scriptsize{Cyrillic\_a} & \scriptsize{} \\
     & U+0410 &  & \scriptsize{Cyrillic\_A} & \scriptsize{} \\
C02 & U+0441 & U+00A9 & \scriptsize{Cyrillic\_es} & \scriptsize{copyright} \\
     & U+0421 & U+00A9 & \scriptsize{Cyrillic\_ES} & \scriptsize{copyright} \\
C03 & U+0434 &  & \scriptsize{Cyrillic\_de} & \scriptsize{} \\
     & U+0414 &  & \scriptsize{Cyrillic\_DE} & \scriptsize{} \\
C04 & U+0444 &  & \scriptsize{Cyrillic\_ef} & \scriptsize{} \\
     & U+0424 &  & \scriptsize{Cyrillic\_EF} & \scriptsize{} \\
C05 & U+0433 &  & \scriptsize{Cyrillic\_ghe} & \scriptsize{} \\
     & U+0413 &  & \scriptsize{Cyrillic\_GHE} & \scriptsize{} \\
C06 & U+0445 &  & \scriptsize{Cyrillic\_ha} & \scriptsize{} \\
     & U+0425 &  & \scriptsize{Cyrillic\_HA} & \scriptsize{} \\
C07 & U+0439 & U+046D & \scriptsize{Cyrillic\_shorti} & \scriptsize{U046D} \\
     & U+0419 & U+046C & \scriptsize{Cyrillic\_SHORTI} & \scriptsize{U046C} \\
C08 & U+043A &  & \scriptsize{Cyrillic\_ka} & \scriptsize{} \\
     & U+041A &  & \scriptsize{Cyrillic\_KA} & \scriptsize{} \\
C09 & U+043B &  & \scriptsize{Cyrillic\_el} & \scriptsize{} \\
     & U+041B &  & \scriptsize{Cyrillic\_EL} & \scriptsize{} \\
C10 & U+003B & U+2026 & \scriptsize{semicolon} & \scriptsize{ellipsis} \\
     & U+003A & U+2026 & \scriptsize{colon} & \scriptsize{ellipsis} \\
C11 & U+0027 & U+2019 & \scriptsize{apostrophe} & \scriptsize{rightsinglequotemark} \\
     & U+0022 & U+2018 & \scriptsize{quotedbl} & \scriptsize{leftsinglequotemark} \\
C12 & U+044E &  & \scriptsize{Cyrillic\_yu} & \scriptsize{} \\
     & U+042E &  & \scriptsize{Cyrillic\_YU} & \scriptsize{} \\
B00 & U+045D &  & \scriptsize{U045D} & \scriptsize{} \\
     & U+040D &  & \scriptsize{U040D} & \scriptsize{} \\
B01 & U+0437 &  & \scriptsize{Cyrillic\_ze} & \scriptsize{} \\
     & U+0417 &  & \scriptsize{Cyrillic\_ZE} & \scriptsize{} \\
B02 & U+044C & U+044B & \scriptsize{Cyrillic\_softsign} & \scriptsize{Cyrillic\_yeru} \\
     & U+045D & U+042B & \scriptsize{U045D} & \scriptsize{Cyrillic\_YERU} \\
B03 & U+0446 & U+00A9 & \scriptsize{Cyrillic\_tse} & \scriptsize{copyright} \\
     & U+0426 & U+00A9 & \scriptsize{Cyrillic\_TSE} & \scriptsize{copyright} \\
B04 & U+0436 &  & \scriptsize{Cyrillic\_zhe} & \scriptsize{} \\
     & U+0416 &  & \scriptsize{Cyrillic\_ZHE} & \scriptsize{} \\
B05 & U+0431 &  & \scriptsize{Cyrillic\_be} & \scriptsize{} \\
     & U+0411 &  & \scriptsize{Cyrillic\_BE} & \scriptsize{} \\
B06 & U+043D &  & \scriptsize{Cyrillic\_en} & \scriptsize{} \\
     & U+041D &  & \scriptsize{Cyrillic\_EN} & \scriptsize{} \\
B07 & U+043C &  & \scriptsize{Cyrillic\_em} & \scriptsize{} \\
     & U+041C &  & \scriptsize{Cyrillic\_EM} & \scriptsize{} \\
B08 & U+002C & U+00AB & \scriptsize{comma} & \scriptsize{guillemotleft} \\
     & U+201E & U+00AB & \scriptsize{doublelowquotemark} & \scriptsize{guillemotleft} \\
B09 & U+002E & U+00BB & \scriptsize{period} & \scriptsize{guillemotright} \\
     & U+201C & U+00BB & \scriptsize{leftdoublequotemark} & \scriptsize{guillemotright} \\
B10 & U+002F & U+0300 & \scriptsize{slash} & \scriptsize{U0300} \\
     & U+003F & U+0301 & \scriptsize{question} & \scriptsize{U0301} \\
интервал & U+0020 & U+00A0 & \scriptsize{space} & \scriptsize{nobreakspace} \\
     & U+0020 & U+00A0 & \scriptsize{space} & \scriptsize{nobreakspace} \\
\end{longtable}

\subsection{Фонетичен вариант за разположението на знаците според БДС 5237:2006} 

\begin{longtable}{lllll}
  \toprule
   & \multicolumn{2}{c}{\tabtit{Unicode}}
   & \multicolumn{2}{c}{\tabtit{X keysym}} \\
  \cmidrule(rl){2-3}
  \cmidrule(l){4-5}
  \tabtit{Клавиш} & \tabtit{1 и 2 рег.} & \tabtit{3 и 4 рег.}
                  & \tabtit{1 и 2 рег.} & \tabtit{3 и 4 рег.} \\
  \midrule
  \endhead
  \bottomrule
  \endfoot
E00 & U+044E &  & \scriptsize{Cyrillic\_yu} & \scriptsize{} \\
     & U+042E &  & \scriptsize{Cyrillic\_YU} & \scriptsize{} \\
E01 & U+0031 &  & \scriptsize{1} & \scriptsize{} \\
     & U+0021 &  & \scriptsize{exclam} & \scriptsize{} \\
E02 & U+0032 &  & \scriptsize{2} & \scriptsize{} \\
     & U+0040 &  & \scriptsize{at} & \scriptsize{} \\
E03 & U+0033 &  & \scriptsize{3} & \scriptsize{} \\
     & U+2116 &  & \scriptsize{numerosign} & \scriptsize{} \\
E04 & U+0034 & U+20AC & \scriptsize{4} & \scriptsize{EuroSign} \\
     & U+0024 & U+20AC & \scriptsize{dollar} & \scriptsize{EuroSign} \\
E05 & U+0035 &  & \scriptsize{5} & \scriptsize{} \\
     & U+0025 &  & \scriptsize{percent} & \scriptsize{} \\
E06 & U+0036 &  & \scriptsize{6} & \scriptsize{} \\
     & U+20AC &  & \scriptsize{EuroSign} & \scriptsize{} \\
E07 & U+0037 &  & \scriptsize{7} & \scriptsize{} \\
     & U+00A7 &  & \scriptsize{section} & \scriptsize{} \\
E08 & U+0038 &  & \scriptsize{8} & \scriptsize{} \\
     & U+002A &  & \scriptsize{asterisk} & \scriptsize{} \\
E09 & U+0039 & U+005B & \scriptsize{9} & \scriptsize{bracketleft} \\
     & U+0028 & U+2329 & \scriptsize{parenleft} & \scriptsize{U2329} \\
E10 & U+0030 & U+005D & \scriptsize{0} & \scriptsize{bracketright} \\
     & U+0029 & U+232A & \scriptsize{parenright} & \scriptsize{U232A} \\
E11 & U+002D & U+2011 & \scriptsize{minus} & \scriptsize{U2011} \\
     & U+2013 & U+2011 & \scriptsize{endash} & \scriptsize{U2011} \\
E12 & U+003D & U+2014 & \scriptsize{equal} & \scriptsize{emdash} \\
     & U+002B & U+2020 & \scriptsize{plus} & \scriptsize{dagger} \\
D01 & U+0447 &  & \scriptsize{Cyrillic\_che} & \scriptsize{} \\
     & U+0427 &  & \scriptsize{Cyrillic\_CHE} & \scriptsize{} \\
D02 & U+0448 &  & \scriptsize{Cyrillic\_sha} & \scriptsize{} \\
     & U+0428 &  & \scriptsize{Cyrillic\_SHA} & \scriptsize{} \\
D03 & U+0435 & U+044D & \scriptsize{Cyrillic\_ie} & \scriptsize{Cyrillic\_e} \\
     & U+0415 & U+042D & \scriptsize{Cyrillic\_IE} & \scriptsize{Cyrillic\_E} \\
D04 & U+0440 & U+00AE & \scriptsize{Cyrillic\_er} & \scriptsize{registered} \\
     & U+0420 & U+00AE & \scriptsize{Cyrillic\_ER} & \scriptsize{registered} \\
D05 & U+0442 & U+2122 & \scriptsize{Cyrillic\_te} & \scriptsize{trademark} \\
     & U+0422 & U+2122 & \scriptsize{Cyrillic\_TE} & \scriptsize{trademark} \\
D06 & U+044A & U+046B & \scriptsize{Cyrillic\_hardsign} & \scriptsize{U046B} \\
     & U+042A & U+046A & \scriptsize{Cyrillic\_HARDSIGN} & \scriptsize{U046A} \\
D07 & U+0443 &  & \scriptsize{Cyrillic\_u} & \scriptsize{} \\
     & U+0423 &  & \scriptsize{Cyrillic\_U} & \scriptsize{} \\
D08 & U+0438 & U+045D & \scriptsize{Cyrillic\_i} & \scriptsize{U045D} \\
     & U+0418 & U+040D & \scriptsize{Cyrillic\_I} & \scriptsize{U040D} \\
D09 & U+043E &  & \scriptsize{Cyrillic\_o} & \scriptsize{} \\
     & U+041E &  & \scriptsize{Cyrillic\_O} & \scriptsize{} \\
D10 & U+043F &  & \scriptsize{Cyrillic\_pe} & \scriptsize{} \\
     & U+041F &  & \scriptsize{Cyrillic\_PE} & \scriptsize{} \\
D11 & U+044F & U+0463 & \scriptsize{Cyrillic\_ya} & \scriptsize{U0463} \\
     & U+042F & U+0462 & \scriptsize{Cyrillic\_YA} & \scriptsize{U0462} \\
D12 & U+0449 &  & \scriptsize{Cyrillic\_shcha} & \scriptsize{} \\
     & U+0429 &  & \scriptsize{Cyrillic\_SHCHA} & \scriptsize{} \\
C01 & U+0430 &  & \scriptsize{Cyrillic\_a} & \scriptsize{} \\
     & U+0410 &  & \scriptsize{Cyrillic\_A} & \scriptsize{} \\
C02 & U+0441 & U+00A9 & \scriptsize{Cyrillic\_es} & \scriptsize{copyright} \\
     & U+0421 & U+00A9 & \scriptsize{Cyrillic\_ES} & \scriptsize{copyright} \\
C03 & U+0434 &  & \scriptsize{Cyrillic\_de} & \scriptsize{} \\
     & U+0414 &  & \scriptsize{Cyrillic\_DE} & \scriptsize{} \\
C04 & U+0444 &  & \scriptsize{Cyrillic\_ef} & \scriptsize{} \\
     & U+0424 &  & \scriptsize{Cyrillic\_EF} & \scriptsize{} \\
C05 & U+0433 &  & \scriptsize{Cyrillic\_ghe} & \scriptsize{} \\
     & U+0413 &  & \scriptsize{Cyrillic\_GHE} & \scriptsize{} \\
C06 & U+0445 &  & \scriptsize{Cyrillic\_ha} & \scriptsize{} \\
     & U+0425 &  & \scriptsize{Cyrillic\_HA} & \scriptsize{} \\
C07 & U+0439 & U+046D & \scriptsize{Cyrillic\_shorti} & \scriptsize{U046D} \\
     & U+0419 & U+046C & \scriptsize{Cyrillic\_SHORTI} & \scriptsize{U046C} \\
C08 & U+043A &  & \scriptsize{Cyrillic\_ka} & \scriptsize{} \\
     & U+041A &  & \scriptsize{Cyrillic\_KA} & \scriptsize{} \\
C09 & U+043B &  & \scriptsize{Cyrillic\_el} & \scriptsize{} \\
     & U+041B &  & \scriptsize{Cyrillic\_EL} & \scriptsize{} \\
C10 & U+003B & U+2026 & \scriptsize{semicolon} & \scriptsize{ellipsis} \\
     & U+003A & U+2026 & \scriptsize{colon} & \scriptsize{ellipsis} \\
C11 & U+0027 & U+2019 & \scriptsize{apostrophe} & \scriptsize{rightsinglequotemark} \\
     & U+0022 & U+2018 & \scriptsize{quotedbl} & \scriptsize{leftsinglequotemark} \\
C12 & U+044C & U+044B & \scriptsize{Cyrillic\_softsign} & \scriptsize{Cyrillic\_yeru} \\
     & U+045D & U+042B & \scriptsize{U045D} & \scriptsize{Cyrillic\_YERU} \\
B00 & U+045D &  & \scriptsize{U045D} & \scriptsize{} \\
     & U+040D &  & \scriptsize{U040D} & \scriptsize{} \\
B01 & U+0437 &  & \scriptsize{Cyrillic\_ze} & \scriptsize{} \\
     & U+0417 &  & \scriptsize{Cyrillic\_ZE} & \scriptsize{} \\
B02 & U+0436 &  & \scriptsize{Cyrillic\_zhe} & \scriptsize{} \\
     & U+0416 &  & \scriptsize{Cyrillic\_ZHE} & \scriptsize{} \\
B03 & U+0446 & U+00A9 & \scriptsize{Cyrillic\_tse} & \scriptsize{copyright} \\
     & U+0426 & U+00A9 & \scriptsize{Cyrillic\_TSE} & \scriptsize{copyright} \\
B04 & U+0432 &  & \scriptsize{Cyrillic\_ve} & \scriptsize{} \\
     & U+0412 &  & \scriptsize{Cyrillic\_VE} & \scriptsize{} \\
B05 & U+0431 &  & \scriptsize{Cyrillic\_be} & \scriptsize{} \\
     & U+0411 &  & \scriptsize{Cyrillic\_BE} & \scriptsize{} \\
B06 & U+043D &  & \scriptsize{Cyrillic\_en} & \scriptsize{} \\
     & U+041D &  & \scriptsize{Cyrillic\_EN} & \scriptsize{} \\
B07 & U+043C &  & \scriptsize{Cyrillic\_em} & \scriptsize{} \\
     & U+041C &  & \scriptsize{Cyrillic\_EM} & \scriptsize{} \\
B08 & U+002C & U+00AB & \scriptsize{comma} & \scriptsize{guillemotleft} \\
     & U+201E & U+00AB & \scriptsize{doublelowquotemark} & \scriptsize{guillemotleft} \\
B09 & U+002E & U+00BB & \scriptsize{period} & \scriptsize{guillemotright} \\
     & U+201C & U+00BB & \scriptsize{leftdoublequotemark} & \scriptsize{guillemotright} \\
B10 & U+002F & U+0300 & \scriptsize{slash} & \scriptsize{U0300} \\
     & U+003F & U+0301 & \scriptsize{question} & \scriptsize{U0301} \\
интервал & U+0020 & U+00A0 & \scriptsize{space} & \scriptsize{nobreakspace} \\
     & U+0020 & U+00A0 & \scriptsize{space} & \scriptsize{nobreakspace} \\
\end{longtable}

\newpage 
\subsection{Разположението на знаците в режим „латиница“}

\begin{longtable}{lllll}
  \toprule
   & \multicolumn{2}{c}{\tabtit{Unicode}}
   & \multicolumn{2}{c}{\tabtit{X keysym}} \\
  \cmidrule(rl){2-3}
  \cmidrule(l){4-5}
  \tabtit{Клавиш} & \tabtit{1 и 2 рег.} & \tabtit{3 и 4 рег.}
                  & \tabtit{1 и 2 рег.} & \tabtit{3 и 4 рег.} \\
  \midrule
  \endhead
  \bottomrule
  \endfoot
E00 & U+0060 & U+2248 & \scriptsize{grave} & \scriptsize{U2248} \\
     & U+007E & U+2245 & \scriptsize{asciitilde} & \scriptsize{U2245} \\
E01 & U+0031 & U+00AC & \scriptsize{1} & \scriptsize{notsign} \\
     & U+0021 & U+00AC & \scriptsize{exclam} & \scriptsize{notsign} \\
E02 & U+0032 & U+00B2 & \scriptsize{2} & \scriptsize{twosuperior} \\
     & U+0040 & U+2022 & \scriptsize{at} & \scriptsize{enfilledcircbullet} \\
E03 & U+0033 & U+00B3 & \scriptsize{3} & \scriptsize{threesuperior} \\
     & U+0023 & U+2260 & \scriptsize{numbersign} & \scriptsize{notequal} \\
E04 & U+0034 & U+20AC & \scriptsize{4} & \scriptsize{EuroSign} \\
     & U+0024 & U+00A3 & \scriptsize{dollar} & \scriptsize{sterling} \\
E05 & U+0035 & U+2030 & \scriptsize{5} & \scriptsize{U2030} \\
     & U+0025 & U+2030 & \scriptsize{percent} & \scriptsize{U2030} \\
E06 & U+0036 & U+2227 & \scriptsize{6} & \scriptsize{logicaland} \\
     & U+005E & U+2228 & \scriptsize{asciicircum} & \scriptsize{logicalor} \\
E07 & U+0037 & U+00A7 & \scriptsize{7} & \scriptsize{section} \\
     & U+0026 & U+00A7 & \scriptsize{ampersand} & \scriptsize{section} \\
E08 & U+0038 & U+221E & \scriptsize{8} & \scriptsize{infinity} \\
     & U+002A & U+00D7 & \scriptsize{asterisk} & \scriptsize{multiply} \\
E09 & U+0039 & U+2202 & \scriptsize{9} & \scriptsize{U2202} \\
     & U+0028 & U+2207 & \scriptsize{parenleft} & \scriptsize{nabla} \\
E10 & U+0030 & U+2300 & \scriptsize{0} & \scriptsize{U2300} \\
     & U+0029 & U+2300 & \scriptsize{parenright} & \scriptsize{U2300} \\
E11 & U+002D & U+2011 & \scriptsize{minus} & \scriptsize{U2011} \\
     & U+005F & U+2013 & \scriptsize{underscore} & \scriptsize{endash} \\
E12 & U+003D & U+2014 & \scriptsize{equal} & \scriptsize{emdash} \\
     & U+002B & U+00B1 & \scriptsize{plus} & \scriptsize{plusminus} \\
D01 & U+0071 & U+0398 & \scriptsize{q} & \scriptsize{Greek\_THETA} \\
     & U+0051 & U+03B8 & \scriptsize{Q} & \scriptsize{Greek\_theta} \\
D02 & U+0077 & U+03A9 & \scriptsize{w} & \scriptsize{Greek\_OMEGA} \\
     & U+0057 & U+03C9 & \scriptsize{W} & \scriptsize{Greek\_omega} \\
D03 & U+0065 & U+2203 & \scriptsize{e} & \scriptsize{U2203} \\
     & U+0045 & U+03B5 & \scriptsize{E} & \scriptsize{Greek\_epsilon} \\
D04 & U+0072 & U+00AE & \scriptsize{r} & \scriptsize{registered} \\
     & U+0052 & U+03C1 & \scriptsize{R} & \scriptsize{Greek\_rho} \\
D05 & U+0074 & U+2122 & \scriptsize{t} & \scriptsize{trademark} \\
     & U+0054 & U+03C4 & \scriptsize{T} & \scriptsize{Greek\_tau} \\
D06 & U+0079 & U+00A5 & \scriptsize{y} & \scriptsize{yen} \\
     & U+0059 & U+03C5 & \scriptsize{Y} & \scriptsize{Greek\_upsilon} \\
D07 & U+0075 & U+2229 & \scriptsize{u} & \scriptsize{intersection} \\
     & U+0055 & U+222A & \scriptsize{U} & \scriptsize{union} \\
D08 & U+0069 & U+222B & \scriptsize{i} & \scriptsize{integral} \\
     & U+0049 & U+03B9 & \scriptsize{I} & \scriptsize{Greek\_iota} \\
D09 & U+006F & U+00B0 & \scriptsize{o} & \scriptsize{degree} \\
     & U+004F & U+222E & \scriptsize{O} & \scriptsize{U222E} \\
D10 & U+0070 & U+03A0 & \scriptsize{p} & \scriptsize{Greek\_PI} \\
     & U+0050 & U+03C0 & \scriptsize{P} & \scriptsize{Greek\_pi} \\
D11 & U+005B & U+2286 & \scriptsize{bracketleft} & \scriptsize{U2286} \\
     & U+007B & U+2282 & \scriptsize{braceleft} & \scriptsize{includedin} \\
D12 & U+005D & U+2208 & \scriptsize{bracketright} & \scriptsize{U2208} \\
     & U+007D & U+2209 & \scriptsize{braceright} & \scriptsize{U2209} \\
C01 & U+0061 & U+2200 & \scriptsize{a} & \scriptsize{U2200} \\
     & U+0041 & U+03B1 & \scriptsize{A} & \scriptsize{Greek\_alpha} \\
C02 & U+0073 & U+03A3 & \scriptsize{s} & \scriptsize{Greek\_SIGMA} \\
     & U+0053 & U+03C3 & \scriptsize{S} & \scriptsize{Greek\_sigma} \\
C03 & U+0064 & U+0394 & \scriptsize{d} & \scriptsize{Greek\_DELTA} \\
     & U+0044 & U+03B4 & \scriptsize{D} & \scriptsize{Greek\_delta} \\
C04 & U+0066 & U+03A6 & \scriptsize{f} & \scriptsize{Greek\_PHI} \\
     & U+0046 & U+03C6 & \scriptsize{F} & \scriptsize{Greek\_phi} \\
C05 & U+0067 & U+0393 & \scriptsize{g} & \scriptsize{Greek\_GAMMA} \\
     & U+0047 & U+03B3 & \scriptsize{G} & \scriptsize{Greek\_gamma} \\
C06 & U+0068 & U+2225 & \scriptsize{h} & \scriptsize{U2225} \\
     & U+0048 & U+03B7 & \scriptsize{H} & \scriptsize{Greek\_eta} \\
C07 & U+006A &  & \scriptsize{j} & \scriptsize{} \\
     & U+004A &  & \scriptsize{J} & \scriptsize{} \\
C08 & U+006B & U+03F0 & \scriptsize{k} & \scriptsize{U03F0} \\
     & U+004B & U+03BA & \scriptsize{K} & \scriptsize{Greek\_kappa} \\
C09 & U+006C & U+039B & \scriptsize{l} & \scriptsize{Greek\_LAMBDA} \\
     & U+004C & U+03BB & \scriptsize{L} & \scriptsize{Greek\_lambda} \\
C10 & U+003B & U+00F7 & \scriptsize{semicolon} & \scriptsize{division} \\
     & U+003A & U+00F7 & \scriptsize{colon} & \scriptsize{division} \\
C11 & U+0027 & U+201C & \scriptsize{apostrophe} & \scriptsize{leftdoublequotemark} \\
     & U+0022 & U+201D & \scriptsize{quotedbl} & \scriptsize{rightdoublequotemark} \\
C12 & U+005C & U+2261 & \scriptsize{backslash} & \scriptsize{identical} \\
     & U+007C & U+22A5 & \scriptsize{bar} & \scriptsize{downtack} \\
B00 & U+003C & U+2266 & \scriptsize{less} & \scriptsize{U2266} \\
     & U+003E & U+2267 & \scriptsize{greater} & \scriptsize{U2267} \\
B01 & U+007A & U+2220 & \scriptsize{z} & \scriptsize{U2220} \\
     & U+005A & U+03B6 & \scriptsize{Z} & \scriptsize{Greek\_zeta} \\
B02 & U+0078 & U+039E & \scriptsize{x} & \scriptsize{Greek\_XI} \\
     & U+0058 & U+03BE & \scriptsize{X} & \scriptsize{Greek\_xi} \\
B03 & U+0063 & U+00A9 & \scriptsize{c} & \scriptsize{copyright} \\
     & U+0043 & U+03C7 & \scriptsize{C} & \scriptsize{Greek\_chi} \\
B04 & U+0076 & U+03A8 & \scriptsize{v} & \scriptsize{Greek\_PSI} \\
     & U+0056 & U+03C8 & \scriptsize{V} & \scriptsize{Greek\_psi} \\
B05 & U+0062 & U+03D1 & \scriptsize{b} & \scriptsize{U03D1} \\
     & U+0042 & U+03B2 & \scriptsize{B} & \scriptsize{Greek\_beta} \\
B06 & U+006E & U+207F & \scriptsize{n} & \scriptsize{U207F} \\
     & U+004E & U+03BD & \scriptsize{N} & \scriptsize{Greek\_nu} \\
B07 & U+006D & U+03BC & \scriptsize{m} & \scriptsize{Greek\_mu} \\
     & U+004D & U+03BC & \scriptsize{M} & \scriptsize{Greek\_mu} \\
B08 & U+002C & U+2266 & \scriptsize{comma} & \scriptsize{U2266} \\
     & U+003C & U+21D4 & \scriptsize{less} & \scriptsize{U21D4} \\
B09 & U+002E & U+2267 & \scriptsize{period} & \scriptsize{U2267} \\
     & U+003E & U+21D2 & \scriptsize{greater} & \scriptsize{U21D2} \\
B10 & U+002F & U+2018 & \scriptsize{slash} & \scriptsize{leftsinglequotemark} \\
     & U+003F & U+2019 & \scriptsize{question} & \scriptsize{rightsinglequotemark} \\
интервал & U+0020 & U+00A0 & \scriptsize{space} & \scriptsize{nobreakspace} \\
     & U+0020 & U+00A0 & \scriptsize{space} & \scriptsize{nobreakspace} \\
\end{longtable}

\section{Относно сигурността}

При въвеждането на пароли потребителят обикновено не вижда знаците,
които въвежда.  В случай, че в паролите е допустимо използването на
кирилски букви, съществува опасността клавиатурната да не генерира
знаците, които потребителят очаква (напр. тип „БДС“ вместо фонетичен
или латиница вместо кирилица).  Така идентификацията на потребителя ще
пропадне без потребителят да бъде уведомен за причината.  Затова
състоянието на клавиатурата („кирилица“ или „латиница“, „БДС“ или
„фонетична“) трябва да се изобразява по някакъв начин на видеомонитора
или чрез използване на светлинни индикатори, вградени в клавиатурата.

В случай, че в паролите не се позволява използването на кирилски
букви, а само само латински букви и други ASCII знаци, програмното
осигуряване би трябвало автоматично да превключи клавиатурата в режим
„латиница“.

\newpage 

\appendix

\section{Илюстрация на клавиатурните подредби}\label{sec:ilustrations}

\subsection{Вариант тип „БДС“ за разположението на знаците върху клавишите}
\subsubsection*{Знаци от първи и втори регистър}

\begin{keyboard}
    
\hline \key{)}& \key{!}& \key{?}& \key{+}& \key{"}& \key{\%}& \key{=}&
\key{:}& \key{/}& \key{--}& \key{№}& \key{\$}& \key{\euro}&
\keyspace{5}{$\leftarrow$} \\

\key{(}& \key{1}& \key{2}& \key{3}& \key{4}& \key{5}& \key{6}&
\key{7}& \key{8}& \key{9}& \key{0}& \key{-}& \key{.}&
\keyspace{5}{\keyname{Backspace}} \\

\hline

\keyspace{4}{$\leftrightarrows$}& \key{ы}& \key{У}& \key{Е}& \key{И}&
\key{Ш}& \key{Щ}& \key{К}& \key{С}& \key{Д}& \key{З}& \key{Ц}&
\key{§}& \keyspace{4}{\keyname{Enter}} \\

\keyspace{4}{\keyname{Tab}}& \key{,}& \key{}& \key{}& \key{}&
\key{}& \key{}& \key{}& \key{}& \key{}& \key{}& \key{}&
\key{;}& \keyspace{4}{$\hookleftarrow$} \\

\cline{1-41}

\keyspace{5}{$\Uparrow$\keyname{Caps}}& \key{\`{и}}& \key{Я}& \key{А}&
\key{О}& \key{Ж}& \key{Г}& \key{Т}& \key{Н}& \key{В}& \key{М}&
\key{Ч}& \key{“}& \keyspace{3}{} \\

\keyspace{5}{\keyname{Lock}}& \key{ь}& \key{}& \key{}&
\key{}& \key{}& \key{}& \key{}& \key{}& \key{}& \key{}&
\key{}& \key{„}& \keyspace{3}{} \\

\hline

\keyspace{4}{$\uparrow$}& \key{\`И} & \key{Ю}& \key{Й}& \key{Ъ}&
\key{Э}& \key{Ф}& \key{Х}& \key{П}& \key{Р}& \key{Л}& \key{Б}&
\keyspace{7}{$\uparrow$} \\

\keyspace{4}{\keyname{Shift}}& \key{}& \key{}& \key{}& \key{}& \key{}&
\key{}& \key{}& \key{}& \key{}& \key{}& \key{}&
\keyspace{7}{\keyname{Shift}} \\

\hline

\end{keyboard}

\subsubsection*{Знаци от трети и четвърти регистър}

С /-/ е означен знакът „свързващо тире“.  Знакът свързващ интервал е
поставен върху клавиша интервал, който не е илюстриран тук.

\begin{keyboard}
    

\hline
\key{]}& \keyn{!}& \keyn{?}& \key{$\dagger$}& \keyn{"}&
  \key{$\rangle$}& \key{"---}& \key{\dots}& \key{\'{~}}& \keyn{--}&
  \keyn{№}& \key{\euro}& \keyn{\euro}&
\keyspace{5}{$\leftarrow$} \\

\key{[}& \keyn{1}& \keyn{2}& \keyn{3}& \keyn{4}&
  \key{$\langle$}& \keyn{6}& \keyn{7}& \key{\`{~}}& \keyn{9}&
  \keyn{0}& \key{/-/}& \keyn{.}&
\keyspace{5}{\keyname{Backspace}} \\

\hline

\keyspace{4}{$\leftrightarrows$}& \key{`}& \keyn{У}& \key{Э}& \key{\`{И}}&
\keyn{Ш}& \keyn{Щ}& \key{\copyright}& \key{\copyright}& \keyn{Д}& \keyn{З}& \keyn{Ц}&
\keyn{§}& \keyspace{4}{\keyname{Enter}} \\

\keyspace{4}{\keyname{Tab}}& \key{'}& \keyn{}& \keyn{}& \keyn{}&
\keyn{}& \keyn{}& \keyn{\copyright}& \keyn{\copyright}& \keyn{}& \keyn{}& \keyn{}&
\keyn{;}& \keyspace{4}{$\hookleftarrow$} \\

\cline{1-41}

\keyspace{5}{$\Uparrow$\keyname{Caps}}& \key{Ы}& \key{\usefont{X2}{cmr}{m}{n}\CYRYAT}& \keyn{А}&
\keyn{О}& \keyn{Ж}& \keyn{Г}& \key{\small{™}}& \keyn{Н}& \keyn{В}& \keyn{М}&
\keyn{Ч}& \key{»}& \keyspace{3}{} \\

\keyspace{5}{\keyname{Lock}}& \keyn{ь}& \keyn{}& \keyn{}&
\keyn{}& \keyn{}& \keyn{}& \keyn{™}& \keyn{}& \keyn{}& \keyn{}&
\keyn{}& \key{«}& \keyspace{3}{} \\

\hline

\keyspace{4}{$\uparrow$}& \keyn{}& \keyn{Ю}& \key{I{\usefont{X2}{cmr}{m}{n}\CYRBYUS}}& \key{\usefont{X2}{cmr}{m}{n}\CYRBYUS}&
\keyn{Э}& \keyn{Ф}& \keyn{Х}& \keyn{П}& \key{®}& \keyn{Л}& \keyn{Б}&
\keyspace{7}{$\uparrow$} \\

\keyspace{4}{\keyname{Shift}}& \keyn{}& \keyn{}& \keyn{}& \keyn{}&
\keyn{}& \keyn{}& \keyn{}& \keyn{}& \keyn{®}& \keyn{}& \keyn{}&
\keyspace{7}{\keyname{Shift}} \\

\hline

\end{keyboard}

\newpage

\subsection{Фонетичен вариант за разположението на знаците върху клавишите}
\subsubsection*{Знаци от първи и втори регистър}

\begin{keyboard}

\hline

\key{Ч}& \key{!}& \key{@}& \key{№}& \key{\$}& \key{\%}& \key{\euro}&
\key{§}& \key{*}& \key{(}& \key{)}& \key{--}& \key{+}&
\keyspace{5}{$\leftarrow$} \\

\key{}& \key{1}& \key{2}& \key{3}& \key{4}& \key{5}& \key{6}&
\key{7}& \key{8}& \key{9}& \key{0}& \key{-}& \key{=}& 
\keyspace{5}{\keyname{Backspace}} \\

\hline

\keyspace{4}{$\leftrightarrows$}& \key{Я}& \key{В}& \key{Е}& \key{Р}&
\key{Т}& \key{Ъ}& \key{У}& \key{И}& \key{О}& \key{П}& \key{Ш}&
\key{Щ}& \keyspace{4}{\keyname{Enter}} \\

\keyspace{4}{\keyname{Tab}}& \key{}& \key{}& \key{}& \key{}&
\key{}& \key{}& \key{}& \key{}& \key{}& \key{}& \key{}&
\key{}& \keyspace{4}{$\hookleftarrow$} \\

\cline{1-41}

\keyspace{5}{$\Uparrow$\keyname{Caps}}& \key{А}& \key{С}& \key{Д}& \key{Ф}&
\key{Г}& \key{Х}& \key{Й}& \key{К}& \key{Л}& \key{:}& \key{"}&
\key{Ю}& \keyspace{3}{} \\

\keyspace{5}{\keyname{Lock}}& \key{}& \key{}& \key{}&
\key{}& \key{}& \key{}& \key{}& \key{}& \key{}& \key{;}&
\key{${}^\prime$}& \key{}& \keyspace{3}{} \\

\hline

\keyspace{4}{$\uparrow$}& \key{\`И}& \key{З}& \key{\`{и}}& \key{Ц}& \key{Ж}&
\key{Б}& \key{Н}& \key{М}& \key{„}& \key{“}& \key{?}&
\keyspace{7}{$\uparrow$} \\

\keyspace{4}{\keyname{Shift}}& \key{}& \key{}& \key{ь}& \key{}&
\key{}& \key{}& \key{}& \key{}& \key{,}& \key{.}&
\key{/}& \keyspace{7}{\keyname{Shift}} \\

\hline

\end{keyboard}

\subsubsection*{Знаци от трети и четвърти регистър}

С /-/ е означен знакът „свързващо тире“.  Знакът свързващ интервал е
поставен върху клавиша интервал, който не е илюстриран тук.

\begin{keyboard}

\hline

\keyn{Ч}& \keyn{!}& \keyn{@}& \keyn{№}& \key{\euro}& \keyn{\%}&
\keyn{\euro}& \keyn{§}& \keyn{*}& \key{$\langle$}& \key{$\rangle$}&
\key{/-/}& \key{$\dagger$}&
\keyspace{5}{$\leftarrow$} \\

\keyn{}& \keyn{1}& \keyn{2}& \keyn{3}& \keyn{4}& \keyn{5}& \keyn{6}&
\keyn{7}& \keyn{8}& \key{[}& \key{]}& \keyn{-}& \key{"---}& 
\keyspace{5}{\keyname{Backspace}} \\

\hline

\keyspace{4}{$\leftrightarrows$}&
\key{{\usefont{X2}{cmr}{m}{n}\CYRYAT}}& \keyn{В}& \key{Э}& \key{®}&
\key{{\small {™}}}& \key{{\usefont{X2}{cmr}{m}{n}\CYRBYUS}}& \keyn{У}&
\key{\`И}& \keyn{О}& \keyn{П}& \keyn{Ш}&
\keyn{Щ}& \keyspace{4}{\keyname{Enter}} \\

\keyspace{4}{\keyname{Tab}}& \keyn{}& \keyn{}& \keyn{}& \keyn{}&
\keyn{}& \keyn{}& \keyn{}& \keyn{}& \keyn{}& \keyn{}& \keyn{}&
\keyn{}& \keyspace{4}{$\hookleftarrow$} \\

\cline{1-41}

\keyspace{5}{$\Uparrow$\keyname{Caps}}& \keyn{А}&
\key{{\usefont{T1}{cmr}{m}{n}\copyright}}& \keyn{Д}& \keyn{Ф}&
\keyn{Г}& \keyn{Х}& \key{I{\usefont{X2}{cmr}{m}{n}\CYRBYUS}}&
\keyn{К}& \keyn{Л}& \key{\dots}& \key{`}& \keyn{Ю}& \keyspace{3}{} \\

\keyspace{5}{\keyname{Lock}}& \keyn{}& \keyn{}& \keyn{}&
\keyn{}& \keyn{}& \keyn{}& \keyn{}& \keyn{}& \keyn{}& \keyn{;}&
\key{'}& \keyn{}& \keyspace{3}{} \\

\hline

\keyspace{4}{$\uparrow$}& \keyn{\`И}& \keyn{З}& \key{Ы}&
\key{{\usefont{T1}{cmr}{m}{n}\copyright}}& \keyn{Ж}& \keyn{Б}&
\keyn{Н}& \keyn{М}& \key{«}& \key{»}& \key{\'{~}}&
\keyspace{7}{$\uparrow$} \\

\keyspace{4}{\keyname{Shift}}& \keyn{}& \keyn{}& \keyn{ь}& \keyn{}&
\keyn{}& \keyn{}& \keyn{}& \keyn{}& \keyn{,}& \keyn{.}&
\key{\`{~}}& \keyspace{7}{\keyname{Shift}} \\

\hline

\end{keyboard}

\newpage

\subsection{Фонетичен вариант за разположението на знаците върху
  клавишите според БДС 5237:2006}
\subsubsection*{Знаци от първи и втори регистър}

\begin{keyboard}

\hline

\key{Ю}& \key{!}& \key{@}& \key{№}& \key{\$}& \key{\%}& \key{\euro}&
\key{§}& \key{*}& \key{(}& \key{)}& \key{--}& \key{+}&
\keyspace{5}{$\leftarrow$} \\

\key{}& \key{1}& \key{2}& \key{3}& \key{4}& \key{5}& \key{6}&
\key{7}& \key{8}& \key{9}& \key{0}& \key{-}& \key{=}& 
\keyspace{5}{\keyname{Backspace}} \\

\hline

\keyspace{4}{$\leftrightarrows$}& \key{Ч}& \key{Ш}& \key{Е}& \key{Р}&
\key{Т}& \key{Ъ}& \key{У}& \key{И}& \key{О}& \key{П}& \key{Я}&
\key{Щ}& \keyspace{4}{\keyname{Enter}} \\

\keyspace{4}{\keyname{Tab}}& \key{}& \key{}& \key{}& \key{}&
\key{}& \key{}& \key{}& \key{}& \key{}& \key{}& \key{}&
\key{}& \keyspace{4}{$\hookleftarrow$} \\

\cline{1-41}

\keyspace{5}{$\Uparrow$\keyname{Caps}}& \key{А}& \key{С}& \key{Д}& \key{Ф}&
\key{Г}& \key{Х}& \key{Й}& \key{К}& \key{Л}& \key{:}& \key{"}&
\key{\`{и}}& \keyspace{3}{} \\

\keyspace{5}{\keyname{Lock}}& \key{}& \key{}& \key{}&
\key{}& \key{}& \key{}& \key{}& \key{}& \key{}& \key{;}&
\key{${}^\prime$}& \key{ь}& \keyspace{3}{} \\

\hline

\keyspace{4}{$\uparrow$}& \key{\`И}& \key{З}& \key{Ж}& \key{Ц}& \key{В}&
\key{Б}& \key{Н}& \key{М}& \key{„}& \key{“}& \key{?}&
\keyspace{7}{$\uparrow$} \\

\keyspace{4}{\keyname{Shift}}& \key{}& \key{}& \key{}& \key{}& \key{}&
\key{}& \key{}& \key{}& \key{,}& \key{.}&
\key{/}& \keyspace{7}{\keyname{Shift}} \\

\hline

\end{keyboard}

\subsubsection*{Знаци от трети и четвърти регистър}

С /-/ е означен знакът „свързващо тире“.  Знакът свързващ интервал е
поставен върху клавиша интервал, който не е илюстриран тук.

\begin{keyboard}


\hline

\keyn{Ю}& \keyn{!}& \keyn{@}& \keyn{№}& \key{\euro}& \keyn{\%}&
\keyn{\euro}& \keyn{§}& \keyn{*}& \key{$\langle$}& \key{$\rangle$}&
\key{/-/}& \key{$\dagger$}&
\keyspace{5}{$\leftarrow$} \\

\keyn{}& \keyn{1}& \keyn{2}& \keyn{3}& \keyn{4}& \keyn{5}& \keyn{6}&
\keyn{7}& \keyn{8}& \key{[}& \key{]}& \keyn{-}& \key{"---}& 
\keyspace{5}{\keyname{Backspace}} \\

\hline

\keyspace{4}{$\leftrightarrows$}&
\keyn{Ч}& \keyn{Ш}& \key{Э}& \key{®}&
\key{{\small {™}}}& \key{{\usefont{X2}{cmr}{m}{n}\CYRBYUS}}& \keyn{У}&
\key{\`И}& \keyn{О}& \keyn{П}& \key{{\usefont{X2}{cmr}{m}{n}\CYRYAT}}&
\keyn{Щ}& \keyspace{4}{\keyname{Enter}} \\

\keyspace{4}{\keyname{Tab}}& \keyn{}& \keyn{}& \keyn{}& \keyn{}&
\keyn{}& \keyn{}& \keyn{}& \keyn{}& \keyn{}& \keyn{}& \keyn{}&
\keyn{}& \keyspace{4}{$\hookleftarrow$} \\

\cline{1-41}

\keyspace{5}{$\Uparrow$\keyname{Caps}}& \keyn{А}&
\key{{\usefont{T1}{cmr}{m}{n}\copyright}}& \keyn{Д}& \keyn{Ф}&
\keyn{Г}& \keyn{Х}& \key{I{\usefont{X2}{cmr}{m}{n}\CYRBYUS}}&
\keyn{К}& \keyn{Л}& \key{\dots}& \key{`}& \key{Ы}& \keyspace{3}{} \\

\keyspace{5}{\keyname{Lock}}& \keyn{}& \keyn{}& \keyn{}&
\keyn{}& \keyn{}& \keyn{}& \keyn{}& \keyn{}& \keyn{}& \keyn{;}&
\key{'}& \keyn{}& \keyspace{3}{} \\

\hline

\keyspace{4}{$\uparrow$}& \keyn{Ж}& \keyn{З}& \keyn{Ы}&
\key{{\usefont{T1}{cmr}{m}{n}\copyright}}& \keyn{В}& \keyn{Б}&
\keyn{Н}& \keyn{М}& \key{«}& \key{»}& \key{\'{~}}&
\keyspace{7}{$\uparrow$} \\

\keyspace{4}{\keyname{Shift}}& \keyn{}& \keyn{}& \keyn{ь}& \keyn{}&
\keyn{}& \keyn{}& \keyn{}& \keyn{}& \keyn{,}& \keyn{.}&
\key{\`{~}}& \keyspace{7}{\keyname{Shift}} \\

\hline

\end{keyboard}

\newpage

\subsection{Разположението на знаците върху клавишите в режим „латиница“}
\subsubsection*{Знаци от първи и втори регистър}

\begin{keyboard}

\hline

\key{$\tilde{}$}& \key{!}& \key{@}& \key{\#}& \key{\$}& \key{\%}&
\key{$\hat{}$}& \key{\&}& \key{*}& \key{(}& \key{)}& \key{\_}&
\key{+}& \keyspace{5}{$\leftarrow$} \\

\key{\`{~}}& \key{1}& \key{2}& \key{3}& \key{4}& \key{5}& \key{6}&
\key{7}& \key{8}& \key{9}& \key{0}& \key{-}& \key{=}& 
\keyspace{5}{\keyname{Backspace}} \\

\hline

\keyspace{4}{$\leftrightarrows$}& \key{Q}& \key{W}& \key{E}& \key{R}&
\key{T}& \key{Y}& \key{U}& \key{I}& \key{O}& \key{P}& \key{\{}&
\key{\}}& \keyspace{4}{\keyname{Enter}} \\

\keyspace{4}{\keyname{Tab}}& \key{}& \key{}& \key{}& \key{}&
\key{}& \key{}& \key{}& \key{}& \key{}& \key{}& \key{[}&
\key{]}& \keyspace{4}{$\hookleftarrow$} \\

\cline{1-41}

\keyspace{5}{$\Uparrow$\keyname{Caps}}& \key{A}& \key{S}& \key{D}& \key{F}&
\key{G}& \key{H}& \key{J}& \key{K}& \key{L}& \key{:}& \key{"}&
\key{|}& \keyspace{3}{} \\

\keyspace{5}{\keyname{Lock}}& \key{}& \key{}& \key{}&
\key{}& \key{}& \key{}& \key{}& \key{}& \key{}& \key{;}&
\key{${}^\prime$}& \key{$\backslash$}& \keyspace{3}{} \\

\hline

\keyspace{4}{$\uparrow$}& \key{>}& \key{Z}& \key{X}& \key{C}& \key{V}&
\key{B}& \key{N}& \key{M}& \key{<}& \key{>}& \key{?}&
\keyspace{7}{$\uparrow$} \\

\keyspace{4}{\keyname{Shift}}& \key{<}& \key{}& \key{}& \key{}& \key{}&
\key{}& \key{}& \key{}& \key{,}& \key{.}&
\key{/}& \keyspace{7}{\keyname{Shift}} \\

\hline

\end{keyboard}

\subsubsection*{Знаци от трети и четвърти регистър}

С /-/ е означен знакът „свързващо тире“.  Знакът свързващ интервал е
поставен върху клавиша интервал, който не е илюстриран тук.

\begin{keyboard}

\hline

\key{$\cong$}& \key{$\lnot$}& \key{$\bullet$}& \key{$\neq$}&
\key{{\usefont{T1}{cmr}{m}{n}\pounds}}&
\key{{\usefont{T1}{cmr}{m}{n}\textperthousand}}& \key{$\vee$}&
\key{§}& \key{$\times$}& \key{$\nabla$}& \keyn{)}& \key{--}&
\key{$\pm$}& \keyspace{5}{$\leftarrow$} \\

\key{$\approx$}& \keyn{1}& \key{${}^2$}& \key{${}^3$}& \key{\euro}&
\keyn{5}& \key{$\wedge$}& \keyn{7}& \key{$\infty$}& \key{$\partial$}&
\key{$\varnothing$}& \key{/-/}& \key{---}&
\keyspace{5}{\keyname{Backspace}} \\

\hline

\keyspace{4}{$\leftrightarrows$}& \key{$\theta$}& \key{$\omega$}&
\key{$\varepsilon$}& \key{$\rho$}& \key{$\tau$}& \key{$\upsilon$}&
\key{$\cup$}& \key{$\iota$}& \key{$\oint$}& \key{$\pi$}&
\key{$\subset$}& \key{$\not\in$}& \keyspace{4}{\keyname{Enter}} \\

\keyspace{4}{\keyname{Tab}}& \key{$\Theta$}& \key{$\Omega$}& \key{$\exists$}&
\key{®}& \key{™}& \key{\yen}& \key{$\cap$}& \key{$\int$}&
\key{\textdegree}& \key{$\Pi$}& \key{$\subseteq$}&
  \key{$\in$}& \keyspace{4}{$\hookleftarrow$} \\

\cline{1-41}

\keyspace{5}{$\Uparrow$\keyname{Caps}}& \key{$\alpha$}& \key{$\sigma$}&
\key{$\delta$}& \key{$\varphi$}& \key{$\gamma$}& \key{$\eta$}& \keyn{J}&
\key{$\kappa$}& \key{$\lambda$}& \key{$\div$}& \key{''}&
\key{$\perp$}& \keyspace{3}{} \\

\keyspace{5}{\keyname{Lock}}& \key{$\forall$}& \key{$\Sigma$}&
\key{$\Delta$}& \key{$\Phi$}& \key{$\Gamma$}& \key{$\Vert$}& \keyn{}&
\key{$\varkappa$}& \key{$\Lambda$}& \keyn{;}& \key{``}&
\key{$\equiv$}& \keyspace{3}{} \\

\hline

\keyspace{4}{$\uparrow$}& \key{$\geqq$}& \key{$\zeta$}& \key{$\xi$}&
\key{$\chi$}& \key{$\psi$}& \key{$\beta$}& \key{$\nu$}& \key{$\mu$}&
\key{$\leftrightarrow$}& \key{$\rightarrow$}& \key{'}&
\keyspace{7}{$\uparrow$} \\

\keyspace{4}{\keyname{Shift}}& \key{$\leqq$}& \key{$\angle$}& \key{$\Xi$}&
\key{©}& \key{$\Psi$}& \key{$\vartheta$}& \key{${}^n$}& \keyn{}& \key{$\leqq$}&
\key{$\geqq$}& \key{`}& \keyspace{7}{\keyname{Shift}} \\

\hline

\end{keyboard}

\newpage






\vspace{20mm}

Адрес на автора:
\begin{quote}
\textit{
Антон Зиновиев \\
Катедра по математическа логика и приложенията \`и \\
Факултет по математика и информатика \\
бул. „Джеймс Баучър“ №5 \\
София 1164 \\
\\
Телефон:  0877 227 400 \\
Ел. поща: anton@lml.bas.bg }
\end{quote}

\end{document}